\documentclass[12pt]{article}
\usepackage[utf8]{inputenc}
\usepackage{pgf}
\usepackage{tikz}
\usepackage[numbers]{natbib}
\usetikzlibrary{positioning}
\usepackage{amsmath}
\usepackage{float}
\usepackage{graphicx}
\usepackage{tikz}
\usetikzlibrary{arrows}
\usepackage{caption}
\usepackage[export]{adjustbox}
\usepackage{lscape} 
\usepackage{multirow}
\usepackage{longtable}
\usepackage{nameref}
\usepackage[margin=1in]{geometry}
\usepackage{url}
\usepackage{dsfont}
\usepackage{makecell}
\usepackage{array}
\usepackage{colortbl}
\usepackage{hhline}
\newcolumntype{L}[1]{>{\raggedright\let\newline\\\arraybackslash\hspace{0pt}}m{#1}}
\newcolumntype{C}[1]{>{\centering\let\newline\\\arraybackslash\hspace{0pt}}m{#1}}
\newcolumntype{R}[1]{>{\raggedleft\let\newline\\\arraybackslash\hspace{0pt}}m{#1}}
\newcolumntype{M}[1]{>{\centering\arraybackslash}m{#1}}
\newcolumntype{N}{@{}m{0pt}@{}}

\title{Just what the doctor ordered: An evaluation of provider preference-based Instrumental Variable methods in observational studies, with application for comparative effectiveness of type 2 diabetes therapy}
\author{Laura M. G\"{u}demann, Beverley M. Shields, John M. Dennis, Jack Bowden\\ on behalf of the MASTERMIND consortium}

\begin{document}
\maketitle

\noindent {\bf Abstract}\\
\\
\noindent Instrumental Variables provide a way of addressing bias due to unmeasured confounding when estimating treatment effects using observational data. As instrument prescription preference of individual healthcare providers has been proposed. Because prescription preference is hard to measure and often unobserved, a surrogate measure constructed from available data is often required for the analysis. Different construction methods for this surrogate measure are possible, such as simple rule-based methods which make use of the observed treatment patterns, or more complex model-based methods that employ formal statistical models to explain the treatment behaviour whilst considering measured confounders. The choice of construction method relies on aspects like data availability within provider, missing data in measured confounders, and possible changes in prescription preference over time.
\\
\\
In this paper we conduct a comprehensive simulation study to evaluate different construction methods for surrogates of  prescription preference under different data conditions, including: different provider sizes, missing covariate data, and change in preference. We also propose a novel model-based construction method to address between provider differences and change in prescription preference. All presented construction methods are exemplified in a case study of the relative glucose lowering effect of two type 2 diabetes treatments in observational data.
\\
\\
Our study shows that preference-based Instrumental Variable methods can be a useful tool for causal inference from observational health data. The choice of construction method should be driven by the data condition at hand. Our proposed method is capable of estimating the causal treatment effect without bias in case of sufficient prescription data per provider, changing prescription preference over time and non-ignorable missingness in measured confounders. 

\section{Introduction}
\begin{figure}[H]
	\centering
	\includegraphics[width=0.4\linewidth]{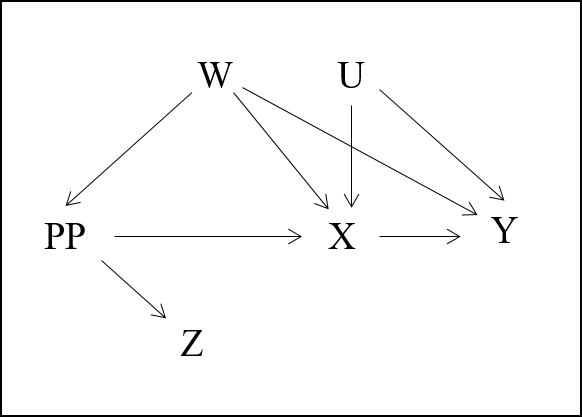}
	\caption{Causal diagram of the proxy design. The true underlying provider preference is not measured but instead approximated with the surrogate variable $Z$.}
	\label{fig:proxy_design_DAG}
\end{figure}

When comparing the relative effectiveness of treatments in observational data,  the Instrumental Variables method (IV)  provides a possibility to address bias in the estimation of treatment effects due to unmeasured confounding. The approach aims to create pseudo-randomized treatment assignment based a suitable instrument for which three main assumptions are necessary. The chosen instrument must strongly predict the treatment decision, be independent of unmeasured confounders and should only affect the outcome through influencing the treatment decision. Korn and Baumrind \cite{Korn98} proposed to utilize healthcare provider prescription preference (PP) as an IV. This instrument has been widely applied in health research areas such as in comparative effectiveness studies for cancer, cardiovascular diseases and mental health. \cite{Widding21}
\\
\\
Preference-based IVs have been constructed at  three different healthcare provider levels in the literature: regional \cite{Bidulka21},  hospitals, practices (or other institutions) \cite{Dalsgaard14}, or at the individual physician level \cite{Davies13a}. In order for PP to be a valid instrument, prescription habits must differ across providers in a manner that cannot be purely explained by patient characteristics which are prognostic for the studied disease, or regional variation in treatment guidelines. Provider preference must also be unrelated to the use of other medical interventions that might influence the outcome. Lastly, it is necessary that patients are assigned to provider independently of their prescription pattern. \cite{Korn98, Brookhart06b}\\
\\
In most routine clinical databases, the reason underlying a treatment decision is not systematically recorded, meaning the true prescription preference of a healthcare provider for one treatment over another is unknown. It is also likely to be a non-binary and evolving variable, representing the strength of a provider's belief in what constitutes the best treatment option for a patient at a particular point in time. Surveys have been designed in order to elicit PP information  \cite{Swanson15,Boef16} but it remains a difficult quantity to accurately and unbiasedly measure \cite{Davies13c}. Therefore, comparative effectiveness studies based on observational data typically do not incorporate data on true provider preference of one drug over the other. In most IV studies using preference-based instruments, PP is instead substituted with a surrogate variable that is somehow estimated from the data. This is referred to as the `proxy design' \cite{Widding21, Hernan06a, Davies17}, which is represented in Figure \ref{fig:proxy_design_DAG}. Here, PP is depicted as valid but unmeasured instrument. It is approximated with the variable $Z$ which utilizes  provider's manifest and observed prescribing behaviour to reflect on PP. The variables $Y$, $W$ and $U$ describe the outcome variable, measured and unmeasured confounders respectively.

\noindent Different methods to construct the instrument have been proposed in the literature. These can be differentiated into two general groups: simple  rule-based methods that utilize subsets of the observed treatment decision data to derive $Z$ at the point a given patient is treated; and more complex methods derived from fitting formal statistical models to the full data (encompassing treatment decisions, outcomes and measured confounders). When multiple surrogate construction methods for $Z$ are available, Brookhart et al. \cite{Brookhart06b} suggest to chose the one that appears to be most strongly related to the observed treatment decision, among those that are unrelated to measured confounders. \cite{Brookhart06b} This is motivated by the desire to minimise the measurement error of $Z$ measuring PP and makes the consideration of additional aspects about prescription preferences necessary. \\
\\
The prescribing preferences of providers undoubtedly contain a dynamic aspect due to the accumulating personal positive or negative experience with administering a drug over time, as well as external factors such as treatment guidelines from health authorities, marketing activities of pharmaceutical companies, and efficacy and safety information from  clinical trials. \cite{Brookhart06b, Abrahamowicz11, Brookhart10, Stafford04, Jackevicius01} Abrahamowicz et al. \cite{Abrahamowicz11} propose a construction method for $Z$ that aims to identify if a provider changes preference and at what point in time the change took place. If such a change point is identified, $Z$ is constructed separately for patients treated before and after the change. \cite{Abrahamowicz11} Clearly, when implementing a particular model-based method, it  is necessary to ensure that  sufficient data per provider is available to support it. \cite{Brookhart06b}\\
\\
Valid IV estimates generally also depend on including treatment effect confounders that are associated with the instrument in the analysis. Missing data are common in observational studies and often dealt with by complete case analysis or applying imputation methods. These strategies can lead to bias in the IV estimates, in case of non-ignorable missingness for which  the missingness depends on  the unmeasured value itself or other unobserved variables. \cite{Little02, Yang14, Ertefaie17} In this challenging setting, Ertefaie et al. \cite{Ertefaie17} propose a model-based  method for constructing $Z$ and show using theoretical arguments and a simulation study that the method is capable of producing unbiased treatment effect estimates.
\\
\\
The aim of this study is a state of the art evaluation of the performance of different construction methods of preference-based instruments with respect to these three data structure aspects and a focus on the more complex model-based approaches. Additionally, we propose an extension of the method by Ertefaie et al. \cite{Ertefaie17} to accommodate non-ignorable missingness as well as possible change in prescription preference.  In Section \ref{sec:ppIV_constructionmethods}  methods for constructing the surrogate variable $Z$ for PP are described, with a focus on the model-based approaches. In Section \ref{sec:simulation_study}, the performance of these methods is evaluated in a simulation study which allows for a change in prescription preference, different data availability and different missingness mechanisms for measured confounders. In Section \ref{sec:application} all construction methods of $Z$ are applied  to primary care data from the Clinical Practice Research Datalink (CPRD) for a comparative effectiveness study comparing  two oral type 2 diabetes (T2D) agents - Sodium-glucose Cotransporter-2 Inhibitors (SGLT2i) versus Dipeptidyl peptidase-4 inhibitors (DPP4i) - on their ability to lower blood glucose levels (HbA1c mmol/mol). Lastly, Section \ref{sec:discussion} concludes the main points of the simulation and application case study and highlights limitations as well as further research possibilities.

\newpage

\section{Constructing preference-based Instrumental \\ Variables}\label{sec:ppIV_constructionmethods}

In this section  we give an overview of the possibilities for constructing a provider preference surrogate variable seen in the literature. We categorize them into two groups based on their use of data on treatment behaviour and measured confounders. Simple rule-based approaches make use of the observed treatment patterns of each provider, while more complex model-based methods use formal statistical models to explain the treatment behaviour, additionally taking the data structure of measured confounders into consideration. The latter category of construction methods focuses on important aspects of preference-based IVs such as possible change in PP over time and the existence of non-ignorable missingness in confounder data. We will focus predominantly on the latter model-based case in the simulation study and application case study of Section \ref{sec:simulation_study}. 

\subsection{Notation}

We assume a study population of $N$ patients, who are clustered into $j=1,...,J$ disjoint sets representing distinct treatment decisions. Provider $j$ treats $i = 1, \ldots, n_j$ patients, so that $N = \sum^{J}_{j=1}n_{j}$. Within each provider, the patients' index $i$ is assumed to coincide with the order in which they have been treated, from first to most recent. The outcome of interest for patient $i$ of provider $j$ is denoted by $Y_{ji}$. Likewise, binary treatment variable $X_{ji}$ denotes whether a patient receives treatment A ($X_{ji} = 0$) or treatment B ($X_{ji} = 1$). Confounders are classified as either measured or unmeasured, and are represented by the $P$- and $M$-length vectors $\mathbf{W_{ji}}$ = ($W_{1ji}, \ldots, W_{Pji}$) and $\mathbf{U_{ji}}$ = ($U_{1ji}, \ldots, U_{Mji}$), respectively. Let the variable $PP_{ji}$ represent the {\it true} underling preference for treatment B over treatment A of provider $j$ at the point they treat patient $i$. We assume that provider preference satisfies the instrumental variable assumptions, either marginally or conditional on $\mathbf{W_{ji}}$.  Finally, let $Z_{ji}$ represent a data-derived {\it estimate} for $PP_{ji}$ which will be used as a proxy measure. In the following sections, the indexes $i$ and $j$ will be omitted from the notation if explanations are not specific to certain providers or individuals.
\\
\\
A visual summary of how each method constructs the IV $Z_{ji}$ within 
a hypothetical provider $j$ is provided in Figure \ref{fig:ppIV_construction_1} for the rule-based methods and in Figure \ref{fig:ppIV_construction_2} for the established model-based approaches. Their respective $x$-axes denote the $i = 1, \ldots, n_j$ patients treated by provider $j$ in sequential order. Prescription decisions $X_{ji}$ are given on the $y$-axes and indicated with a cross symbol. The true underlying provider preference $PP_{ji}$ is indicated by the solid line. The corresponding derived instrument is marked with a circle. Shaded areas indicate which treatment data is utilised in the derivation of the respective $Z_{ji}$.   It also clarifies for which patients $Z_{ji}$ cannot be calculated.\\
\\
Constructing $Z$ is a means to an end, the end being to perform an IV analysis to overcome confounding between observed values of $X$ and $Y$. Throughout this paper we will implement this analysis within the classic Two-Stage Least Squares framework by doing the following:
\begin{enumerate}
    \item{Use data $\mathbf{W}$ and $X$ to derive instrument $Z$};
\item{Regress: $X$ on $Z$ and $\mathbf{W}$ to obtain a predicted value $\hat{X}$};
\item{Regress: $Y$ on  $\hat{X}$ and $\mathbf{W}$ to obtain an estimate for the causal effect of receiving treatment A ($X$ = 0) versus B ($X = 1$) on the outcome}
\end{enumerate}
The focus of this section is on different methods for performing step 1.

 \subsection{Rule-based approaches}\label{sec:ppIV_rulebased}
\begin{figure}[H]
	\centering
	\includegraphics[width=0.8\linewidth]{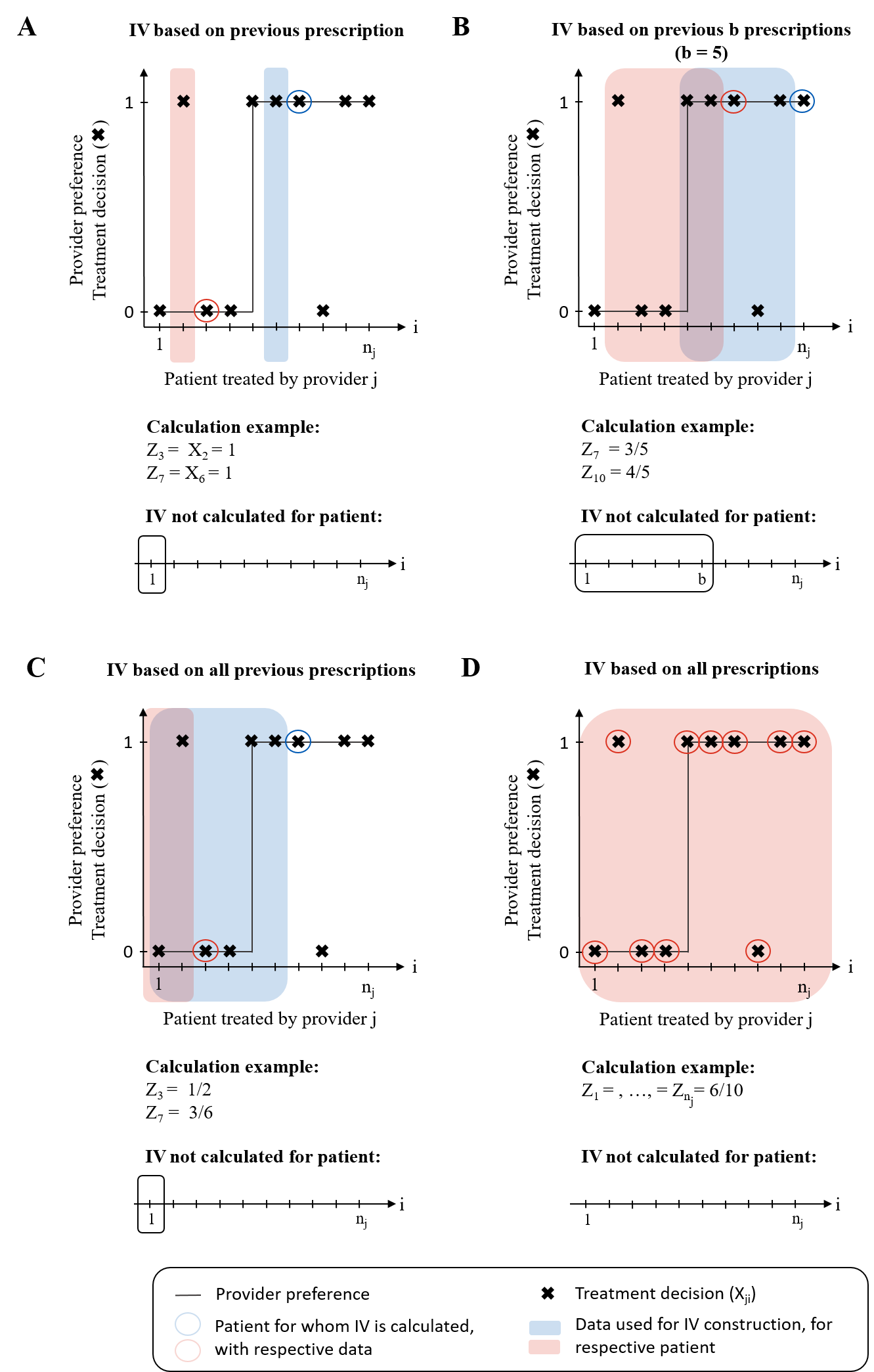}
	\caption{{Visualization of the rule-based preference-based IV construction methods with corresponding calculation example. Abbreviations used for the methods are A: IV prevpatient, B: IV prevbpatient, C: IV allprevprop, D: IV allprop, IV alldichmean, IV alldichmedian.}}
	\label{fig:ppIV_construction_1}
\end{figure}

Brookhart et al. \cite{Brookhart06b} proposed to conceptualize $Z_{ji}$ with the most recent prescription. That is, for each patient $i$ and treated by provider $j$,  $Z_{ji} = X_{ji-1}$. Therefore, $Z_{ji}$  is a binary variable and is calculable for all patients treated by  provider $j$ {\it except} the first one. As $Z_{ji}$ is constructed using the most recent prior treatment decision, it can reflect on true changes in provider preference but also random fluctuations from patient to patient not indicative of a genuine change or trend \cite{Brookhart06b, Abrahamowicz11, Rassen09b}. This method will be referred to as {\it IV prevpatient} and is visualized in panel A of Figure \ref{fig:ppIV_construction_1}. \\
\\
A generalisation of this construction method calculates $Z_{ji}$ as the proportion of patients who received a given treatment, say treatment B, among the previous $b$ treated patients, so that 

\[\ 
Z_{ji} = \frac{1}{b}\sum^{i-1}_{d=i-(b-1)}X_{jd} \quad = \quad b\bar{X}_{i}(b).
\]
This instrument can take on values  in the range of $0\%$ and $100\%$ as opposed to being a binary variable. \cite{Rassen09b, Hennessy08, Uddin16} We refer to this method as {\it IV prevbpatient} by setting $b$ in the following sections equal to 2, 5, and 10. This method can theoretically reflect on changes in the providers preference depending on how close to the time of preference change patient $i$ is treated, but is based on more data. Clearly $Z_{ji}$ cannot be calculated for the first $b$ patients within each provider, which means that some data is lost. A calculation example for $b = 5$ is given in panel B of Figure \ref{fig:ppIV_construction_1}. \\
\\
A closely related variation of this method is to construct $Z_{ji}$ using {\it all} previous prescriptions for treatment $B$, so that:
\[\ 
Z_{ji} = \frac{1}{i-1}\sum^{i-1}_{d=1}X_{jd} \quad = \quad (i-1)\bar{X}_{i}(i-1).
\]
It can be calculated for all patients (except the first) and will be referred to as {\it IV allprevprop}. \cite{Brookhart06b} This method is summarized in panel C of Figure \ref{fig:ppIV_construction_1}.
\\
\\
For the previous methods $Z_{ji}$ is calculated individually for each patient $i$ within provider $j$. Alternatively, all prescription data can be utilized to derive a single instrument for the provider. This leads to 

\[\ 
Z_{ji} = \frac{1}{n_{j}}\sum^{n_{j}}_{i=1}X_{ji} \quad = \quad n_{j}\bar{X}_{j} = Z_{j},
\]

\noindent which can be calculated for $i = 1, \ldots, n_j$  and lies between 0\% and 100\%. This will be referred to as {\it IV allprop}. It is possible to dichotomize $Z_{ji}$ and created a binary instrument with the median \cite{Uddin16} or mean empirical value of all practitioners, so that in the case of the median:

\begin{equation}
    Z_{ji}=
    \begin{cases}
      0, & \text{if}\ n_{j}\bar{X}_{j} \leq \text{Median}(n_{1}\bar{X}_{1},\ldots,n_{J}\bar{X}_{J}) \\
      1, & \text{otherwise.} \nonumber
    \end{cases}
  \end{equation}
These methods will be referred to as, {\it IV alldichmedian} and {\it IV alldichmean} respectively. A minimum provider size $n_{j,min} = 2$ is needed to apply this method and its dichotomized versions. The methods are represented in panel D of Figure \ref{fig:ppIV_construction_1}. \\
\\
For the estimation of the treatment effect with any given preference-based IV, the data of patients for which $Z_{ji}$ cannot be calculated are excluded. This has clear implications for the efficiency of each method.

\subsection{Model-based approaches}

We now introduce two model-based approaches to construct $Z$: the method of Ertefaie et al. \cite{Ertefaie17} (henceforth the `Ertefaie' method) which fits a multi-level model and is capable of dealing with non-ignorable missingness in a measured confounder; and secondly, the method of Abrahamowicz et al. \cite{Abrahamowicz11} (henceforth the `Abrahamowicz' method) which tests for and then potentially models a change point in provider preference over time. We additionally propose an extension of the Ertefaie method that allows for time trend in provider preference using a mixed model with a random intercept and slope. 

\subsubsection{The Ertefaie method}\label{sec:ppIV_Ertefaie}
\begin{figure}[H]
	\centering
	\includegraphics[width=0.8\linewidth]{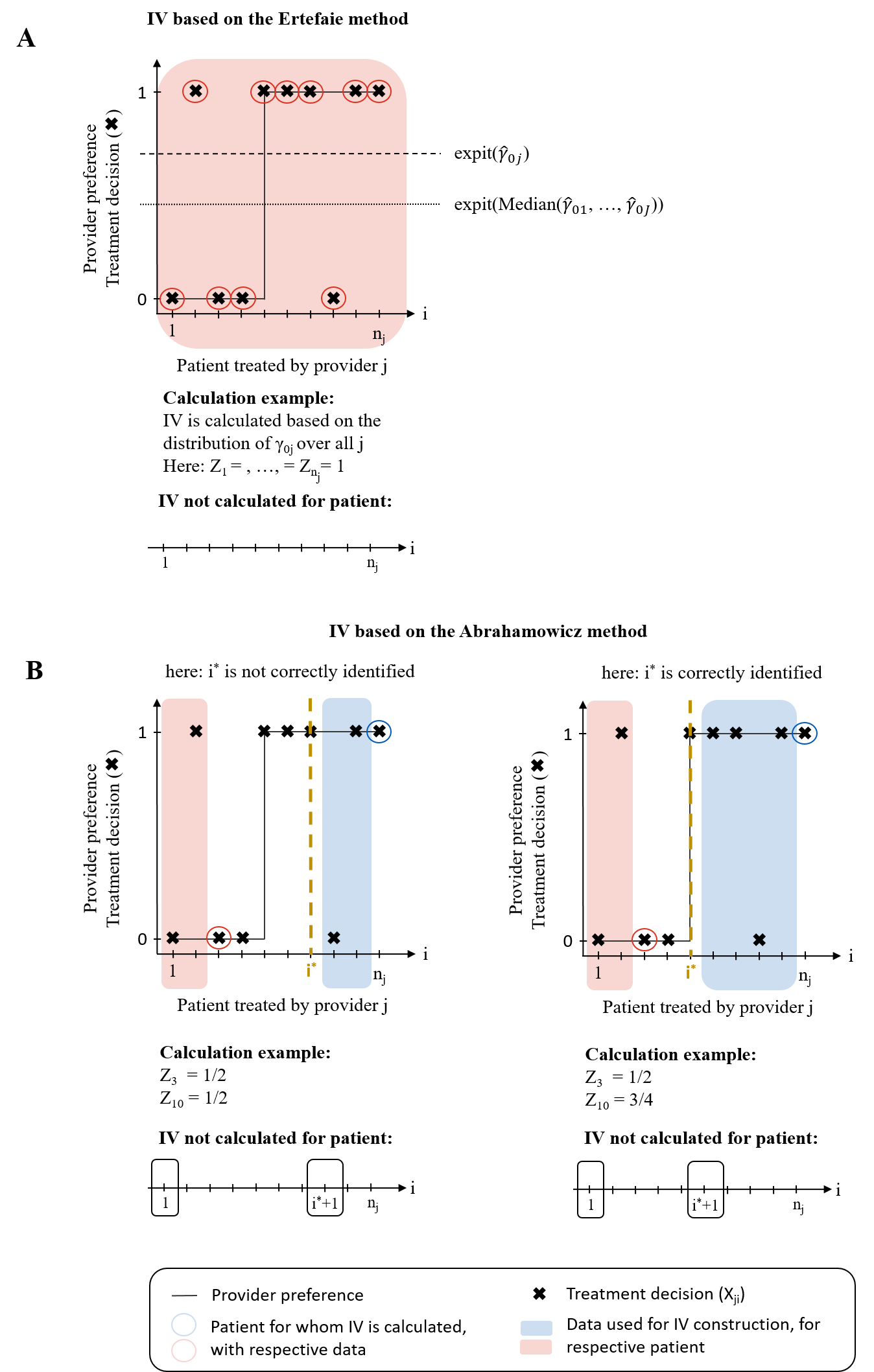}
	\caption{{Visualization of the established model-based preference-based IV construction method by Ertefaie et al. 2017 and Abrahamowics et al. 2011. The Abrahamowicz method is shown with correctly identified (left side) and incorrectly identified (right side) change time $i^\star$. Abbreviations used for the methods are A: IV ePP, B: IV star.}}
	\label{fig:ppIV_construction_2}
\end{figure}

Ertefaie et al. \cite{Ertefaie17} propose a procedure for the estimation of a valid treatment effect utilizing IVs based on provider preference in the presence of baseline characteristics with non-ignorable missingness. For this approach the set of measured baseline characteristics is subdivided in $\mathbf{W_{obs,ji}}$ for all confounders fully observed, and $\mathbf{W_{miss,ji}}$ denoting all confounders which are not completely recorded for all $i$. 
\\
\\
The instrument $Z_{ji}$ is estimated from a generalized random multilevel model, regressing $X_{ji}$ on $\mathbf{W_{obs,ji}}$ and $\mathbf{W_{miss,ji}}$. The model includes the random intercept $\gamma_{0j}$ for each provider (provID) and is estimated using a complete case dataset on all measured confounders: 
\begin{equation}\label{equ:Ertefaie}
\begin{split}
    Logit(P(X_{ji} = 1| \mathbf{W_{obs,ji}}, \mathbf{W_{miss,ji}}, provID_{ji})) = \gamma_0 + \gamma_{0j} + \gamma_{W_{miss,ji}}\mathbf{W_{miss,ji}} + \\ \gamma_{W_{obs}}\mathbf{W_{obs,ji}} + \varepsilon_{ji}.
\end{split} 
\end{equation}
Ertefaie et al. \cite{Ertefaie17} propose to estimate the instrument for provider $j$ based on the relative position of random intercept $\hat{\gamma}_{0j}$ in the entire distribution, so that: 

\begin{equation}
    Z_{ji} = Z_j =
    \begin{cases}
      1, & \text{if}\left( expit(\hat{\gamma}_{0j}) > expit(\text{Median}\left\{\hat{\gamma}_{01},\ldots,\hat{\gamma}_{0J}\right\}) \right) \\
      0, & \text{otherwise.} \nonumber
    \end{cases}
  \end{equation}
The visualisation of this approach in Figure \ref{fig:ppIV_construction_2} (top) shows  that each patient within provider $j$ will receive the same value for $Z_{ji}$ and that the instrument can be estimated for all $i$ in $j$. The minimum sample size for this construction method can not be precisely determined from data availability as for the rule-based methods, but depends on the estimation of the mixed effect model. An often cited rule of thumb for minimum sample size is Kreft’s 30/30 rule which requires data on at least 30 provider  and 30 patients within each provider. \cite{Kreft96} Further discussion on this can be found for example in Snijders and Bosker \cite{Snijders11} or Hox and van de Schoot \cite{Hook10}.  In Section \ref{sec:simulation_study} the effects of different provider sizes on the estimation performance will be discussed and we employ this construction method as well as its extension method if a provider treats at least two patients. 
\\
\\
An advantage of the  Ertefaie method is that, despite the IV derivation being based on complete case data, the final analysis is employed on the full data, only using  $\mathbf{W_{obs}}$ in the outcome model. The only proviso is that a random intercept and hence $Z_{ji}$ can be derived for a given provider.  If in fact there are no complete case individuals in a given provider, the provider must be excluded from the analysis.
\\
\\
The Ertefaie method relies on  three assumptions. The first assumptions requires missingness to occur at provider level unrelated to unmeasured confounders and the treatment and at individual patient level, which can be dependent on measured and unmeasured confounders. This assumption is plausible for studies in which missingness varies across providers for example due to staff or management policies. In a sensitivity analysis the authors show that the approach is still valid under moderate violations of this assumption \cite{Ertefaie17}. The second assumption states that the effect of $U$ on $X$ should not vary by provider. As this is not testable with the data at hand, the authors suggest that it is possible to make an educated guess by checking whether measured confounders violate this assumption. This could be checked with  a generalized mixed-effect model to predict the treatment decision including measured confounders and provID as random intercept and random slope. Significant random slopes would indicate that this assumption might be violated. The third assumption (positivity) indicates that each provider has a non-zero probability in theory of seeing patients with any given observed characteristics.  The authors verify this approach in a simulation study with non-ignorable missingness in one measured confounder. The treatment effect estimation results are  unbiased even in case of non-ignorable missingness and have lower standard errors compared to standard IV approaches using complete case or multiple imputed datasets. A sensitivity analysis confirms that the approach can even deal with  `high rates' of non-ignorable missingness. \cite{Ertefaie17}  Since the method makes use of a complex mixed effect model and the prescription data over all providers simultaneously, it is hard to relate the derivation of $Z$ to the observed treatment data $X$ in any given provider. Nevertheless, the method is summarized in panel A of Figure \ref{fig:ppIV_construction_2}. When summarizing results in Section \ref{sec:simulation_study} and Section \ref{sec:application} it will be referred to as \textit{IV ePP}.

\subsubsection{The Abrahamowics method }\label{sec:ppIV_method_Abrahamowics}
In order to address concerns about variance inflation when using only one previous prescription to construct $Z$, Abrahamowicz et al. \cite{Abrahamowicz11} propose a more complex modification of IV prevpatient that aims to detect and account for a change in provider preference over time, yet uses more prescription data to achieve smaller estimation variance. The procedure is applied for each of the $J$ providers individually in the following 4 step procedure.  We follow the previous convention by assuming patients within  provider $j$ are ranked by calendar time  from 1 to $n_{j}$.
\\
\\
\noindent \textbf{Step 1:} 
To test if a provider changes preference, a reference no-change model is estimated from a multivariable logistic regression of $X_{ji}$ with $p = 1, \ldots, P$ measured covariates. 
\begin{equation*}
    Logit\left( Pr[X_{ji} = 1)] \right) = \beta_0 + \sum_{p=1}^P \beta_p W_{pji}.  
\end{equation*}
From this model the deviance $D(0)_j$ for provider $j$ is extracted. 
\\
\noindent \textbf{Step 2:} 
This step is applied to test for a change in preference of provider $j$ and is further split up  into three iterations. 
\\
\noindent \textbf{Step 2.1:}
The difference in prescription proportions for each patient $i = 3, \ldots, n_j - 3$ is calculated by subtracting the proportion of $X_{ji} = 1$ of the  `later patients' with $k > i$ from the proportion of `earlier patients' with $k \le i$:
\begin{equation*}
    d_{ji} = \underbrace{\frac{\sum_{1 \le k \le i}}{i\textcolor{white}{_j}}}_\text{earlier patients} - \underbrace{\frac{\sum_{i+1 \le k \le n_j}}{n_j - i}}_\text{later patients} .
\end{equation*}
As this difference in proportions is calculated from the third to the third last patient treated by provider $j$, the minimum number of patients needed within each provider is $5$. The provider preference is assumed to be constant across the observed time period if $|d_{ji} < 0.2|$ $\forall i$ in $i = 3, \ldots, n_j - 3$.  For all other provider with at least one $|d_{ji} \ge 0.2|$ further investigations are conducted to identify possible change in preference. In the next steps the time of the preference change is identified.   
\\
\noindent \textbf{Step 2.2:}
A change-time model is estimated for each patient $i$ with $|d_{ji} \ge 0.2|$:  
\begin{equation*}
    Logit\left( Pr[X_{ji} = 1] \right) = \beta_0 + \sum_{p=1}^P \beta_p W_{pji} + \eta \mathds{1}(k > i) . 
\end{equation*}
In addition to the no-change model, this model includes  a binary variable indicating patients prescribed after $i$ with $\mathds{1}(k > i) = 0$ if $k\le i$ and $\mathds{1}(k > i) = 1$ if $k > i$. Note that for the most recently prescribed patient $i = n_j$,  $\mathds{1}(k > i) = 0$. The approach will therefore not be able to test for a change in preference  directly after the last prescribed patient of provider $j$. The parameter $\eta$ is the average adjusted difference in the propensity of provider $j$ to prescribe $X_{ji} = 1$ between `earlier patients' and `later patients'. From the change-time models the deviances $D_j(i)$ are extracted. 
\\
\noindent \textbf{Step 2.3:} 
In order to identify after which prescription a possible change in preference took place, the optimal change-time model and the time of change $i^\star$ is defined as 
\begin{equation*}
    min(D_j(i) = D_j(i^\star)).
\end{equation*}
\\
\noindent \textbf{Step 3:} 
In this step the fits of the no-change model and the optimal  change-time model are compared using the Akaike information criterion, which is calculated as $AIC = $ deviance $ + 2L$, $L$ being the number of parameters estimated in the model. For the change-time model two additional parameter are estimated, $\eta$ and the optimal change time $i^\star$. Therefore, the no-change model is only considered to have a better fit if
\begin{equation}\label{equ:Abrahamowics_step4}
    D(0)_j < D_j(i^\star) + 4 .
\end{equation}
\\
\noindent \textbf{Step 4:} 
Providers are identified to {\it not} have a change in preference if, as explained in step 2.1 $|d_{ji} \ge 0.2|$ $\forall i$ in $j$ or if the no-change model is the best fit explained in step 3. In this case $Z_{ji}$ for all patients in provider $j$ are constructed as IV allprevprop explained in Section \ref{sec:ppIV_rulebased}. Providers {\it are} identified to have a change in preference if the conditions in step 2.1 and step 2.3 are satisfied. It is then assumed that the change in preference takes place between $i^\star$ and $i^\star + 1$. In this case $Z_{ji}$ is constructed by firstly subdividing the prescription data into two subgroups, before and after the change time. The two groups will encompass patients $i = 1, \ldots i^\star$ and patients $i = i^\star + 1, \ldots n_j$ respectively. Following this, $Z_{ji}$ is calculated as IV allprevprop in both subgroups limited to the treatment information from those subgroups.\\
\\
Panel B of Figure \ref{fig:ppIV_construction_2}  shows a calculation example for $Z_{ji}$, in two cases where  $i^\star$ is correctly  and incorrectly identified. The calculation of $Z_{ji}$ for all patients after the change will depend on the identification of  $i^\star$. Additionally, the graph points out that for providers who have a change identified by the method, $Z_{ji}$ cannot be calculated for the two patients $i = 1$ and $i = i^\star + 1$. Lastly, it should be noted that the identification of a change in preference  relies on the assumption that provider will only change their preference once within the observed time period. When summarizing results in consecutive sections the method will be referred to as \textit{IV star}.

\subsubsection{Extension of the Ertefaie method}\label{sec:ppIV_Ertefaie_rirs}
 We propose an extension to the Ertefaie method, \textit{IV ePP (rirs)},  that utilizes a random intercept random slope model to construct $Z$. This approach aims to combine the original strategy for  estimating unbiased treatment effects even under non-ignorable missingness for measured confounders and the principle idea underpinning the Abrahamowicz method to  account for the possibility of changing prescription preference over the observed period, albeit implemented in a more straightforward model. Specifically, the instrument $Z_{ji}$ is constructed from a generalized random intercept, random slope model for the treatment decision: 
\begin{equation}\label{equ:Ertefaie_rirs}
\begin{split}
    Logit(Pr[X_{ji} = 1| \mathbf{W_{obs,ji}}, \mathbf{W_{miss,ji}}, provID_{ji}, T_{ji}]) = \gamma_0 + \gamma_{0j} + (\gamma_{Tj} + \gamma_{T})T_{ji} + \\ \gamma_{W_{miss}}\mathbf{W_{miss,ji}} + \gamma_{W_{obs}}\mathbf{W_{obs,ji}} + \varepsilon_{ji},
\end{split}
\end{equation}
where $T_{ji}$ represents a variable indicating the time of prescription and $\mathbf{W_{obs,ji}}$, $\mathbf{W_{miss,ji}}$ and $provID_{ji}$ are defined as before. From this model the following fitted values $\hat{\Theta}_{ji}$ are derived with the estimated global intercept $\hat{\gamma}_0$, random intercept $\hat{\gamma}_{0j}$, the global slope for $T_{ji}$ $\hat{\gamma}_T$ and the respective random slope $\hat{\gamma}_{Tj}$: 
\begin{equation}\label{equ:fitted_ePP(rirs)}
	\hat{\Theta}_{ji} = \hat{\gamma}_0 + \hat{\gamma}_{0j} + (\hat{\gamma}_T + \hat{\gamma}_{Tj})T_{ji} .
\end{equation}
Finally, the instrument is constructed by applying the following binary transformation to $\hat{\Theta}_{ji}$:

\begin{equation}
    Z_{ji}=
    \begin{cases}
      1, & \text{if}\left( expit(\hat{\Theta}_{ji}) > expit(\text{Median}\left\{\hat{\Theta}_{11},\ldots,\hat{\Theta}_{Jn_{J}},\right\}) \right) \\
      0, & \text{otherwise.} \nonumber
    \end{cases}
  \end{equation}


\newpage
\section{Simulation study}\label{sec:simulation_study}
Several simulation studies investigating different aspects of provider preference-based IVs can be found in the literature. Ionescu-Ittu et al. \cite{Ionescu09} assesses the performance of using prescription data of previous patients under varying instrument strengths. Abrahamowicz et al.  \cite{Abrahamowicz11} and Ertefaie et al.  \cite{Ertefaie17} also include simulations studies showcasing the key feature of their proposed methods, namely modelling a change in prescription preference and accounting for non-ignorable missingness. 
\\
\\
Our simulation study aims to compare all rule-based and model-based methods for PP IV construction, but with a specific focus on the two model-based approaches proposed by  Abrahamowicz et al. \cite{Abrahamowicz11} and  Ertefaie et al. \cite{Ertefaie17}, as well as our extension method. Results of the rule-based method are referred to and are discussed in more detail in Appendix 2. 

\subsection{Data generation}

Population data is generated for $J = 100$ providers who treat each $1, \ldots, n_j$ patients in ascending order. Two measured covariates $W_{1,ji} \sim N(\mu_{W_1}, 2)$ and $W_{2,ji} \sim N(\mu_{W_2}, 2)$ with $\mu_{W_1}$ and $\mu_{W_2} \sim N(0,0.5)$ and one unmeasured confounder $U_{ji} \sim N(0, 1)$ are simulated.  The outcome of interest, $Y$, is simulated as a continuous variable with 
\begin{equation}
	Y_{ji} = \gamma_{Y, 0} + \beta X_{ji} + \gamma_{Y,W_1}W_{1,ji} + \gamma_{Y,W_2}W_{2,ji} + \gamma_{Y,U}U_{ji} + \varepsilon_{Y, ji}
\end{equation}
and the true (or causal) treatment effect is fixed at $\beta = 1$.\\
\\
In order to not intentionally bias simulation results in favour of either model-based method, treatment decision data $X$ were generated under two separate processes to be congenial for the Abrahamowicz or Ertefaie methods respectively.
 
\subsubsection{Generating X under the Abrahamowicz model} \label{sec:simulation_datageneration}
Following Abrahamowicz et al. \cite{Abrahamowicz11} we simulate the prescription preference of each provider for a given patient using a binary variable $PP_{ji}$ which indicates the preference for treatment B ($X_{ji} = 1$) if $PP_{ji} = 1$. Specifically, our data generating model for $X_{ji}$ is: 
\begin{equation}\label{equ:X_simA}
	X_{ji} \sim Bern(\gamma_{X,0} + \beta_{PP, ji}PP_{ji} + \gamma_{X,U}U_{ji} + \gamma_{X,W_1}W_{1,ji} + \gamma_{X,W_2}W_{2,ji}). 
\end{equation} 
In order to induce a degree of stochasticity into the variable for  $PP$, we use the following procedure:

\begin{itemize}
    \item {The initial preference of provider $j$ is simulated with  $PP_{initial,ji}\sim Bern(0.6)$, so that $40\%$ and 60\%  prefer treatment A and B respectively}
    \item{Original `A preferers' change to `B preferers' with a probability of $70\%$ for some $2\leq i^{*} \leq n_{j}$}
    \item{Original `B preferers' change to `A preferers' with a probability of $40\%$ for some $2\leq i^{*} \leq n_{j}$}
\end{itemize}

\noindent The change time is simulated with $i^{*} \sim U(0.4\times n_j, 0.7\times n_j)$ $\forall j$ and $\beta_{PP,ji} = 0.7$. This means that, on average around $57\%$ of providers will change their preference. 

\subsection{Generating X under the extended Ertefaie model}
 Ertefaie et al. \cite{Ertefaie17} use a mixed effect model with a random intercept to model provider preference. To compliment this we simulate treatment decision data $X_{ji}$ using a mixed effect model analogous to equation \ref{equ:Ertefaie_rirs}, including both a random intercept to model different initial preference levels and a random slope to allow for a change in PP:
\begin{equation}\label{equ:X_simB}
	X_{ji} \sim Bern\left( \gamma_{X,0} + \gamma_{X,0j} + (\gamma_{X,T} + \gamma_{X,Tj})T_{ji} + \gamma_{X,U}U _{ji}+ \gamma_{X,W_1}W_{1,ji} + \gamma_{X,W_2}W_{2,ji}  \right).
\end{equation}
Here, $T_{ji}$ increases from $1$ to  $12$ in ascending order of $i$, and one could view these time points as successive calendar months. The random intercept and random slope parameters are simulated with 
\[\
\begin{bmatrix} \gamma_{X,0j} \\ \gamma_{X,Tj} \end{bmatrix}  \sim N (0, \Omega) \text{  and  } \Omega = \begin{bmatrix}
\sigma^2_{\gamma_{0j}} & \sigma_{\gamma_{T0j}} \\
	\sigma_{\gamma_{0Tj}} & \sigma^2_{\gamma_{Tj}} \end{bmatrix}. \nonumber 
\]

\noindent The simulation was conduced using R version 4.2.1 and the analysis of each scenario was repeated in $200$ simulation runs. R code for the simulation can be found in \url{https://github.com/GuedemannLaura/ ppIV }.  Results are given for both ways of modelling the treatment decision in equation's (\ref{equ:X_simA}) and (\ref{equ:X_simB}).  
\\
\\
Table \ref{tab:SimAandB_additionalinfomation} in Appendix 1 summarises additional information on both data generation strategies. For both strategies the variances of $Y_{ji}$ are similar with $Var(Y_{ji}) \approx 7.7$ for the simulation generating $X_{ji}$ under the Abrahamowicz model and $Var(Y_{ji}) \approx 6.9$ for the simulation using the extended Ertefaie model. Though, the proportion of treated patients ($X_{ji} = 1$)  shows more differences with around $42 \%$ for the former and $56\%$ for the latter simulation strategy.

\subsection{Scenarios}
Two simulation scenarios are chosen to probe different challenges for the analysis: Change in provider preference over time, a variable amount of available prescription data per provider, and missing data in the measured confounders. For the first scenario, the number of treated patients per provider is chosen with $n_j = 24, 108, 408$. The second scenario involves the simulation of missing values in the measured confounder variable $W_{1}$ due to different missing mechanisms: either missing completely at random (MCAR) or a non-ignorable missing data mechanism. The latter will be referred to as a missing not at random mechanism (MNAR). Missingness in $W_{1,ji}$ is indicated with the indicator variable $R_{ji}$ and both mechanism results in $p(R_{ji} = 1) \approx 40\% $.  
\\
\\
For the MNAR mechanism the missingness is simulated using the same data generation process as Ertefaie et al. \cite{Ertefaie17}. The mechanism depends on all measured and unmeasured confounders, the outcome variable and $V$, the provider level influence on missing data. Specifically,  the missingness indicator $ R_{MNAR, ji} \sim Bern(\pi_{R,ji})$, where
\begin{equation} 
	\begin{split}
		 	\pi_{R,ji} = & \frac{exp(\gamma_{R,0} + \gamma_{R,W_1}W_{1,ji} + \gamma_{R,W_2}W_{2,ji} + \gamma_{R,U}U_{ji} + \gamma_{R,Y^{\star}_{ji}} Y^{\star}_{ji})}{1 + exp(\gamma_{R,0} + \gamma_{R,W_1}W_{1,ji} + \gamma_{R,W_2}W_{2,ji} + \gamma_{R,U}U_{ji} + \gamma_{R,Y^{\star}_{ji}} Y^{\star}_{ji})} \\
		&\times \frac{exp(\gamma_{R,0} + \gamma_{R,V}V_{ji} + \gamma_{R,VW_1}V_{ji}W_{1,ji} + \gamma_{R,VW_2}V_{ji}W_{2,ji})}{1 + exp(\gamma_{R,0} + \gamma_{R,V}V_{ji} + \gamma_{R,VW_1}V_{ji}W_{1,ji} + \gamma_{R,VW_2}V_{ji}W_{2,ji})} 
	\end{split}
\end{equation}
with $Y^{\star}$ denoting the standardized outcome variable and $V_{ji}\sim U(-2,2)$. Both scenarios are implemented with change in preference as described above. Table \ref{tab:summary_scenario} summarizes all scenarios and clarifies how they are implemented for each data generation strategy. The focus of each scenario with a change in parameters is indicated in \textbf{bold}.  

\begin{table}[H]
	\begin{tabular}{l|l|l|l}
		\hline
		\multicolumn{2}{l|}{ \textcolor{white}{w}}   & \multicolumn{2}{c}{ \textbf{Data generation of $X$}}                                                          \\ \cline{3-4}
	\multicolumn{2}{l|}{ \textcolor{white}{w}}     & \textbf{Abrahamowicz model}    &\textbf{Extended Ertefaie model}                                                         \\ \hline
		Scenario 1 & \begin{tabular}[c]{@{}l@{}}$n_j$\\ missing data\\ $\beta_{PP,ji}$\\ change in $PP_{ji}$\end{tabular} & \multicolumn{1}{l|}{\begin{tabular}[c]{@{}l@{}}\textbf{24, 108, 408}\\ no NAs\\ $0.7$  $\forall j$\\ some $j$\end{tabular}} & \begin{tabular}[c]{@{}l@{}}\textbf{24, 108, 408}\\ no NAs\\ \\ \\ \\ \end{tabular} \\ \hline
		Scenario 2 & \begin{tabular}[c]{@{}l@{}}$n_j$\\ missing data\\ $\beta_{PP,ji}$\\ change in $PP_{ji}$\end{tabular} & \multicolumn{1}{l|}{\begin{tabular}[c]{@{}l@{}}408\\ \textbf{no NAs, MCAR, MNAR}\\ $0.7$ $\forall j$\\ some $j$\end{tabular}}   & \begin{tabular}[c]{@{}l@{}} 408\\ \textbf{no NAs, MCAR, MNAR} \\ \\ \\ \\ \end{tabular}  \\                                              
	\end{tabular}
\caption{\label{tab:summary_scenario}Summary of the simulation scenarios for the two data generation strategies of $X$. The focus of each scenario and the corresponding change in parameters is highlighted in \textbf{bold}.}
\end{table}

\subsection{Results}
\begin{figure}[H]
	\hspace{-0.3cm}\includegraphics[width=1.04\linewidth, left]{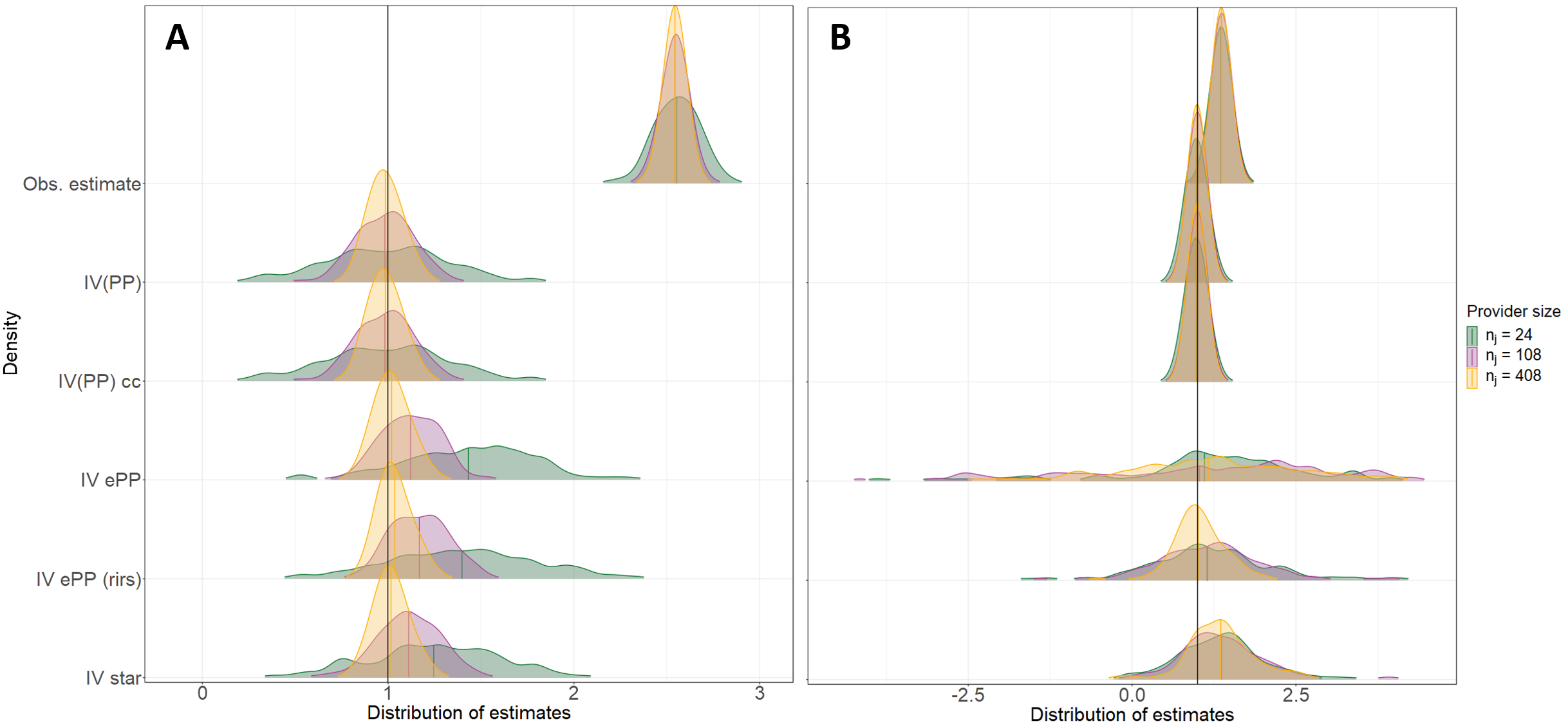}
	\caption{Estimation results of scenario 1: change in provider size. Panel A: estimation results for the data generation process using the Abrahamowicz model. Panel B: estimation results for the data generation process using the extended Ertefaie model to simulate change in preference. Results are summarized for the model-based construction methods of $Z$.}
	\label{fig:simulation_results_scenario1_modelbasedmethods}
\end{figure}

\begin{figure}[H]
	\hspace{-0.3cm}\includegraphics[width=1.04\linewidth, left]{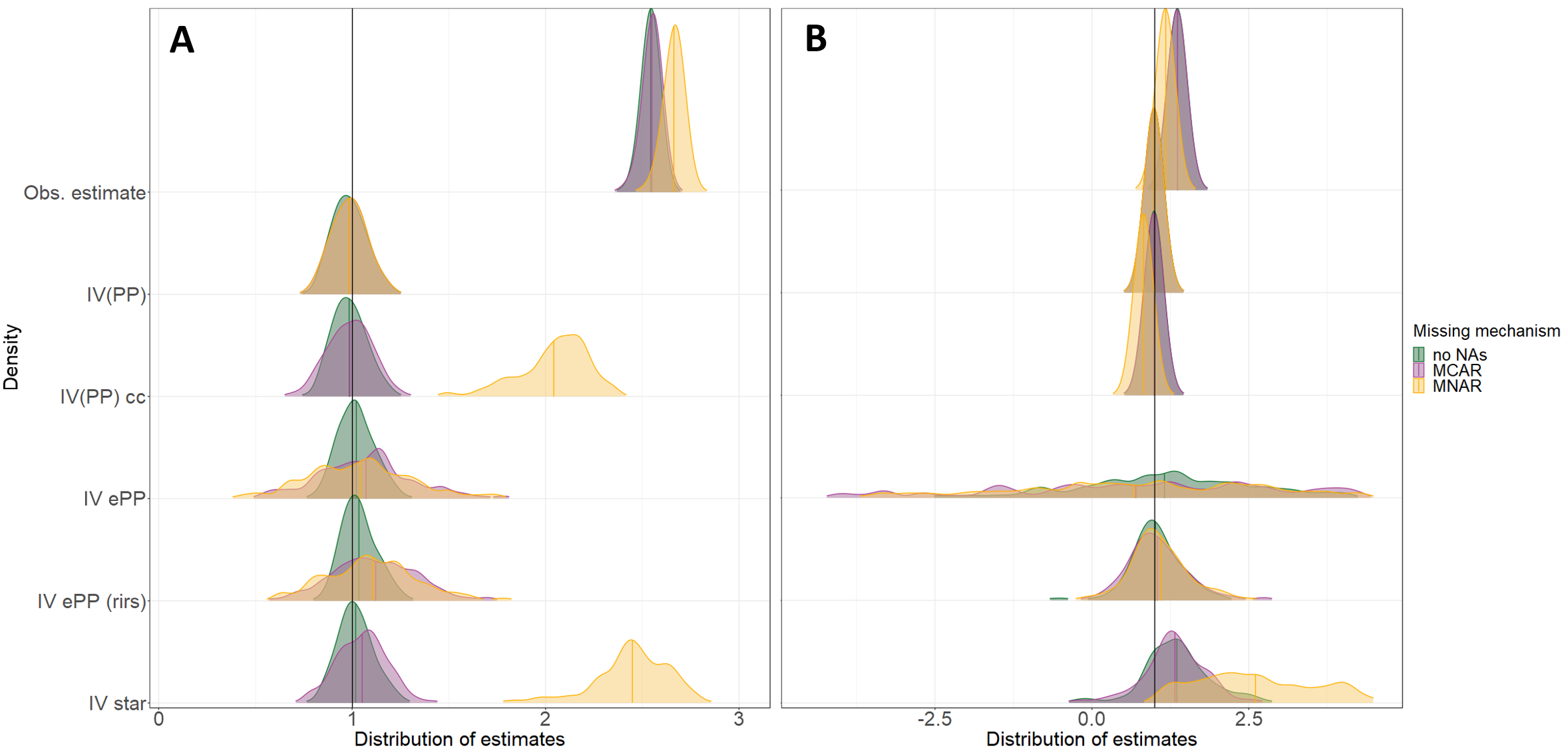}
	\caption{Estimation results of scenario 2: missing data mechanisms. Panel A: estimation results for the data generation process using the Abrahamowicz model. Panel B: estimation results for the data generation process using the extended Ertefaie model to simulate change in preference. Results are summarized for the model-based construction methods of $Z$.}
	\label{fig:simulation_results_scenario2_modelbasedmethods}
\end{figure}

The estimation results of the treatment effect are represented with density plots in Figure \ref{fig:simulation_results_scenario1_modelbasedmethods} and \ref{fig:simulation_results_scenario2_modelbasedmethods} for the model-based methods. Results of the rule-based methods can be found in Appendix 2. As useful benchmark results for the IV estimation, three additional estimates are reported
\begin{itemize}
    \item An `as treated' estimate is calculated  using a multivariable regression model for $Y$ on the observed treatment decision variable $X$ adjusted for the measured confounders only;
    \item IV(PP) describes the estimate using the true simulated $PP$ as IV. Depending on the data generation process the $PP$ variable is either explicitly simulated (Abrahamowicz model) or can be derived with the true value of $\Theta$ calculated with formula \eqref{equ:fitted_ePP(rirs)} (extended Ertefaie model) and the true simulated values for global and random intercept and slope; 
    \item IV(PP) cc is the complete case analysis version of IV(PP).
    
\end{itemize}
As the latter two estimates use the true $PP$, their results provide useful upper bounds for the performance of any method that constructs a provider prescription preference proxy from the data. For all scenarios with NAs, the results of IV(PP) cc give additional insight in the bias caused by applying a complete case analysis. The `as treated' estimate is denoted as \textit{observational estimate} in all result summaries. Density plots will give insights to the bias and estimation variance across different construction methods. Further results on the coverage and relative root mean squared error (RMSE) are given in Appendix 3 for all methods. Besides the estimation performance, Table \ref{tab:f_stats_results_simulationA_modelbased} and \ref{tab:f_stats_results_simulationB_modelbased} summarize the F-statistic of the first-stage regression model from the Two-Stage Least Squares IV estimation and for the model-based construction methods. \cite{Lousdal18} For the rule-based construction methods the F-statistic tables can be found in Appendix 2.  This information is valuable when judging the strength of the instrument as a result from the different  construction methods. Often, the instrument is considered to be a weak instrument in case of F-statistic values smaller than 10. \cite{Stock02} Table \ref{tab:summary_methods_1} and \ref{tab:summary_methods_2} summarize the methods applied in the simulation study and their abbreviations. 

\begin{table}[H]
\begin{tabular}{l|p{12.5cm}}
\hline
Abbreviation     & Method  \\ \hline
Obs. estimate    & Observational estimate, multivariable regression adjusted for \newline measured confounders \\
IV(PP)          & True simulated PP as IV, utilizing all data in case of  missingness                 \\
IV(PP) cc       & True simulated PP as IV, utilizing complete case data in case of  \newline missingness       \\
IV ePP           & IV constructed with the Ertefaie method                                            \\
IV ePP (rirs)    & IV constructed with our proposed extended Ertefaie method                      \\
IV star          & IV constructed with the Abrahamowicz method \\ \hline                             
\end{tabular}
\caption{\label{tab:summary_methods_1} Summary of the model-based IV construction methods and benchmarking methods applied in the simulation and their abbreviations.}
\end{table}

\noindent With scenario 1 the first aspect, different amount of available data, is investigated. The provider sizes are chosen with $n_j = 24, 108, 408$ and estimation results of this scenario are summarized in Figure \ref{fig:simulation_results_scenario1_modelbasedmethods}. Panel A shows the estimation results of the model-based construction methods for the simulation strategy of $X$ simulating $PP$. Estimation of the treatment effect is more efficient with larger $n_j$ and biased for smaller provider sizes. This indicates that we need sufficient data for the more complex models to adequately recover the true $PP$. As IV(PP) is unbiased for all $n_j$, the treatment effect estimates are very likely biased due to measurement error when constructing $Z$ as surrogate for PP.  F-statistic results summarized in Table \ref{tab:f_stats_results_simulationA_modelbased} show that all methods lead to strong instruments. Additionally, Table \ref{tab:performance_measures_scenario1_simulationA} in Appendix 3 shows that all model-based methods show good coverage in case of sufficient provider size (i.e. $n_j = 108$ and $n_j = 408$). \\
\\
From Panel B of Figure \ref{fig:simulation_results_scenario1_modelbasedmethods} it is noticeable that the treatment effect is generally estimated with higher estimation variance when the treatment decision is simulated with a mixed effect model. Only IV ePP (rirs) estimates the treatment effect without bias, given sufficient provider size. IV ePP exhibits only small bias, but also largest estimation variance, as the model of the first step uses only a random intercept to model $X$. This results underlines the importance of specifying the model for $PP$ correctly. All F-statistic results show that the constructed instrument are strong with values larger than 10.

\noindent With scenario 2 the effect on the estimation results due to different mechanism of missing data is analysed. Comparing the estimation results of IV(PP) and IV(PP) cc gives an impression of the magnitude of bias that is caused by MNAR versus no NAs and MCAR when applying a complete case analysis, as done by most of the construction methods. For both data generation strategies the results make clear that only IV ePP and IV ePP (rirs) are able to deal with non-ignorable missingness. All other methods exhibit bias.  Even though, the estimation results of IV star are clearly biased, the F-statistic results indicate that this construction method will lead to a strong instrument. This is a meaningful illustration why testing the IV assumption and choosing a strong IV should not be the only consideration for this application and how important it is to investigate missingness in the data at hand.

\begin{table}[H]
	\begin{center}
		\begin{normalsize}
			\color{black}
			\begin{tabular}{r|rrc|rrc}
				\hline
				 &\multicolumn{3}{c|}{Scenario 1}&\multicolumn{3}{c}{Scenario 2}\\ 
				&\multicolumn{3}{c|}{Provider size ($n_j$)}&\multicolumn{3}{c}{Missing mechanism}\\ 
				&$24$&$108$&$408$&no NAs&MCAR&MNAR\\ 
	\hline
IV(PP) & 27.5 & 116.94 & 451.4 &   451.4 & 275.46 & 238.62 \\ 
IV(PP) cc & 27.5 & 116.94 & 451.4 &   451.4 & 275.46 & 238.62 \\ 
IV ePP & 70.98 & 94.56 & 195.37 &   195.37 & 112.87 & 83.84 \\ 
IV ePP (rirs)& 60.5 & 155.12 & 410.79  & 410.79 & 243.22 & 186.92 \\ 
IV star & 25.2 & 68.51 & 127.99  & 127.99 & 103.04 & 104.31 \\ 
\hline
			\end{tabular}
		\end{normalsize}
	\end{center}
\caption{\label{tab:f_stats_results_simulationA_modelbased} F-statistic results for the instrument $Z$ from the first stage regression model of the Two-Stage Least Squares approach. The results are summarized for all scenarios. For this simulation the treatment decision $X$ with the Abrahamowicz model. This table summarizes the results of all model-based construction methods for $Z$.}
\end{table}
\color{black}


\begin{table}[H]
	\begin{center}
		\begin{normalsize}
			\color{black}
			\begin{tabular}{r|rrc|rrc}
				\hline
				 &\multicolumn{3}{c|}{Scenario 1}&\multicolumn{3}{c}{Scenario 2}\\ 
				&\multicolumn{3}{c|}{Provider size ($n_j$)}&\multicolumn{3}{c}{Missing mechanism}\\ 
				&$24$&$108$&$408$&no NAs&MCAR&MNAR\\ 
	\hline
IV(PP) & 705.35 & 3181.47 & 12031.13 &   12031.13 & 7232.37 & 6790.5 \\ 
IV(PP) cc & 705.35 & 3181.47 & 12031.13 &   12031.13 & 7232.37 & 6790.5 \\ 
IV ePP & 65.53 & 22.41 & 20.64 &   20.64 & 6.24 & 3.13   \\ 
IV ePP (rirs)& 37.28 & 67.2 & 208.57  & 208.57 & 216.5 & 216.61 \\ 
IV star & 48.95 & 111.16 & 147.16  & 147.16 & 129.61 & 374.26 \\  
\hline
			\end{tabular}
		\end{normalsize}
	\end{center}
\caption{\label{tab:f_stats_results_simulationB_modelbased} F-statistic results for the instrument $Z$ from the first stage regression model of the Two-Stage Least Squares approach. The results are summarized for all scenarios. For this simulation the treatment decision $X$ is generated using the extended Ertefaie method. This table summarizes the results of all model-based construction methods for $Z$.}
\end{table}
\color{black}

\noindent With regards to the data availability, results of this simulation study show that the  model-based approaches needed sufficient large amount of prescription data to estimate the treatment effect without bias. For smaller provider sizes, the rule-based methods were capable of estimating the treatment effect without bias. These results are discussed in more detail in Appendix 2. Only the construction method by the Ertefaie method and our extension method were able to adequately estimate the treatment effect in case of MNAR. Whereby, IV ePP showed only small bias but larger estimation variance compared to IV ePP (rirs). All model-based methods where able to produce treatment effects with small bias and acceptable estimation variance for most of the scenarios with a change in preference. The rule-based methods which use all prescription data within a provider to construct $Z$ did struggle to estimate the treatment effect as they do not reflect on the change in $PP$, especially for small provider sizes. Additionally, the type of change did not seem to make much of a difference for the estimation performance for all construction methods. \\
\\
As Uddin et al. \cite{Uddin16} and our own simulation study has concluded, the validity of a preference-based IV strongly depends on the suitability of the data it is applied to. If possible, treatment effect estimates from multiple constructions should be derived and their coherence assessed \cite{Uddin16}. In keeping with this spirit, in Section \ref{sec:application} we apply all of the rule and model-based construction methods discussed thus far to look at the comparative efficiency of two oral type 2 diabetes treatments.

\section{Applied analysis: Comparative effectiveness assessment of two treatments for type 2 diabetes}\label{sec:application}
\begin{figure}[H]
	\centering
	\includegraphics[width=0.9\linewidth]{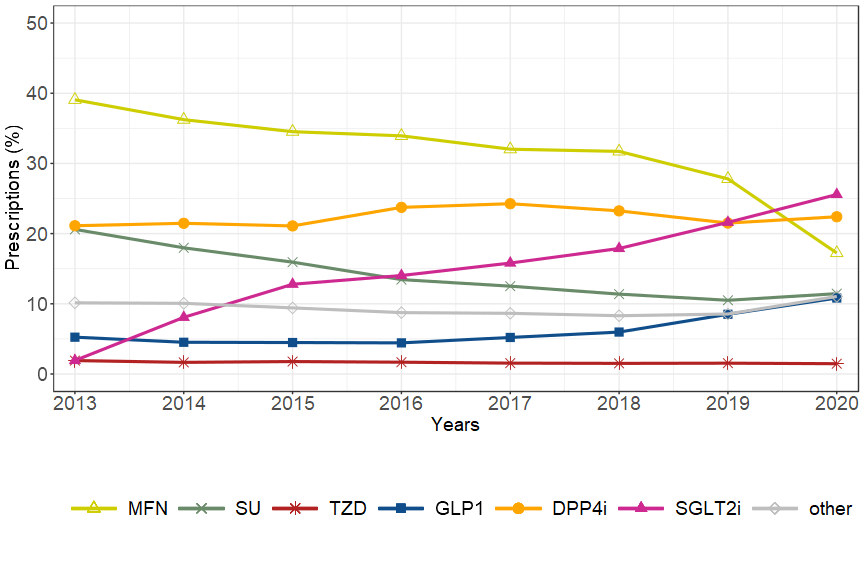}
	\caption{Prescribing trend of T2D oral agents in the study population data in the years 2013 to 2020. The trends are described by the yearly percentage of prescription of each agent respectively and relative to all T2D oral agent prescriptions.}
	\label{fig:prescription_time_trend}
\end{figure}

\begin{figure}[H]
	\centering
	\includegraphics[width=1\linewidth]{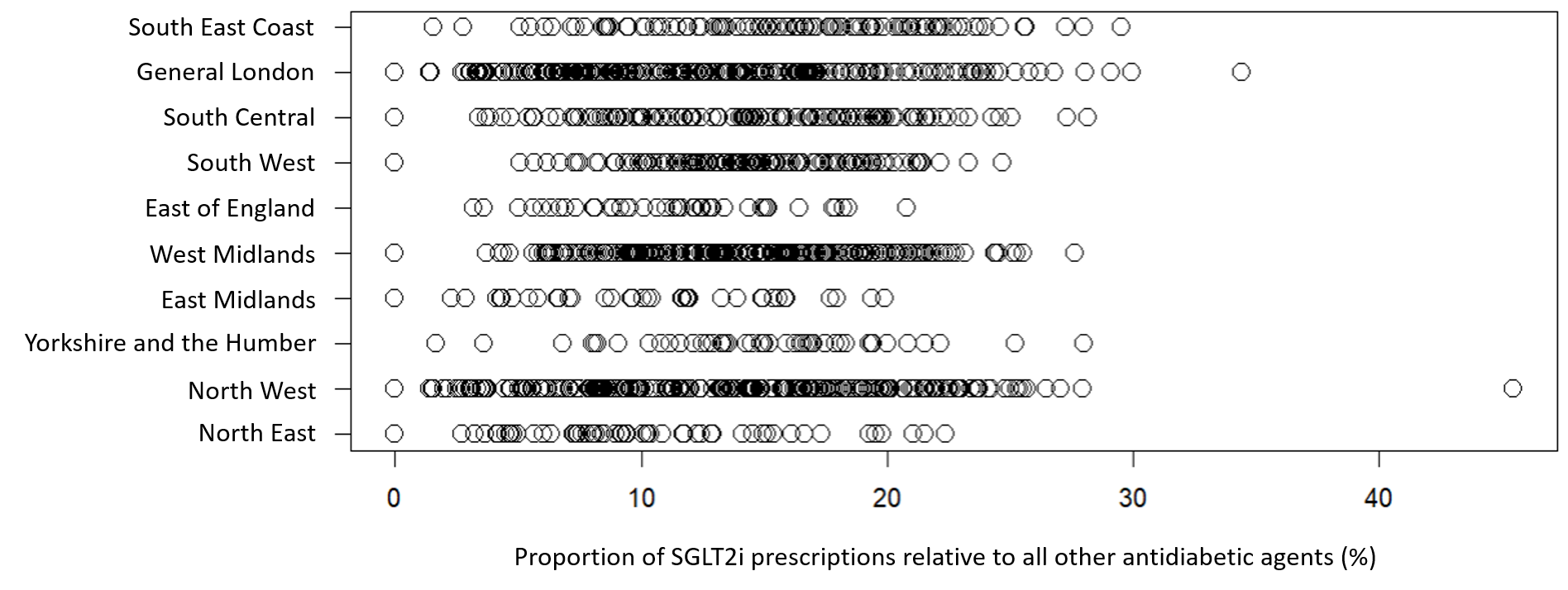}
	\caption{Prescription variation of SGLT2i between all practices in the study population. Each circle represents the proportion of SGLT2i relative to all prescriptions of T2D oral agents within a practice. The practices are clustered within their corresponding region for readability of the plot only.}
	\label{fig:SGLT2i_prescription_within_provider_region}
\end{figure}

Type 2 diabetes (T2D) is a serious progressive metabolic disorder, characterized by hyperglycaemia and with an inherent risk of micro- and macrovascular complications. \cite{Zaccardi16b} Treatment mainly focuses on the control of blood glucose measured by the maintenance of glycated haemoglobin (HbA1c) levels. \cite{Sharma16, Curtis18} HbA1c level management is controlled by lifestyle changes and glucose-lowering agents. Two increasingly prescribed glucose-lowering agents are Sodium-glucose Cotransporter-2 Inhibitors (SGLT2i) and Dipeptidyl peptidase-4 inhibitors. \cite{Curtis18, Dennis19b} Although head-to-head RCT data suggest that the average glucose-lowering efficacy of both therapies is approximately similar \cite{Scheen13}, estimates are derived from highly selected cohorts which are not representative of the wider T2D population. This case study aimed to apply different IV methods in a real-world comparative effectiveness evaluation of the glucose-lowering efficacy of both therapies in an unselected T2D population. For this analysis different IVs for the preference of prescribing SGLT2i over DPP4i were constructed and applied. IV construction took place at the level of the GP practice in UK routine clinical data. UK routine clinical data contain key features of direct relevance to the methods discussed thus far, namely: (1) substantial between practice variation in preference for each drug even when patient covariates are taken into account;  (2) evidence of a trend in prescription preference in favour of SGLT2i over time; and (3) missing data in key patient covariates that one would ideally like to adjust for. Figure  \ref{fig:prescription_time_trend} and \ref{fig:SGLT2i_prescription_within_provider_region} contain further information about the prescription trends between 2013-2020 and the prescription variation of SGLT2i between providers. Figure \ref{fig:prescription_time_trend} highlights the increased prescribing of SGLT2i in recent years, which likely reflects their greater prominence in T2D treatment guidelines due to an accumulation of evidence on their cardiorenal benefits. Each circle in Figure \ref{fig:SGLT2i_prescription_within_provider_region} represents the proportion of SGLT2i prescriptions within a provider, relative to all other T2D oral agent prescriptions. Prescriptions of SGLT2i vary greatly between providers. As prescription of T2D agents is mostly done by primary care practices in the UK and not by specialised practices, a clustering of patients with specific characteristics is unlikely. This gives us confidence that a preference-based IV can be applied with this data. \cite{Korn98, Brookhart06b} A similar study on the cardiovascular safety profile of Sulfonylureas  using provider prescription preference with primary care data from Scotland was conduced by Wang et al. \cite{Wang23}. For this analysis $Z$ was constructed using IV prevbpatient with $b = 10$ or all prescriptions of the previous 365 days.

\subsection{Study Population and Data preparation}
We used routine data from the Clinical Practice Research Datalink (CPRD) Aurum (download November 2021). \cite{Herrett15} The study cohort of identified people with T2D included patients who initiated either SGLT2i ($n = 77229$) or DPP4i ($n = 109608$) between 2013 and 2020. A protocol on the identification strategy of people with T2D is explained in Rodgers et al. \cite{Rodgers17}. CPRD is a large  source of primary healthcare data and encompasses $6.9\%$  of the population in the UK. Furthermore, it is considered to be representative of the UK population regarding age, sex and ethnicity. \cite{Herrett15} The study cohort comprised people with T2D initiating either SGLT2i ($n = 77229$) or DPP4i ($n = 109608$) between 2013 and 2020 who had baseline HbA1c 53-120 mmol/mol and had an estimated glomerular filtration rate (eGFR) of $\ge 45 mL/min/1.73m^2$. The chosen HbA1c range represents the lower threshold for glucose-lowering medication initiation in clinical guidelines and for severe hyperglycemia. SGLT2i’s were contraindicated in the UK in individuals with eGFR $< 45 mL/min/1.73 m^2$ and not licensed for use below this threshold for the majority of the study period. CPRD data extraction followed our previously published protocol \cite{Rodgers17}. Baseline clinical characteristics (measured confounders) are reported in Table \ref{tab:cohort_description} for each drug arm, and were included in the outcome model of 12 month achieved HbA1c (mmol/mol), closest value to 12 months in the 9-15 months after treatment initiation, on unchanged therapy. Each practice treated on average $132$ individuals with either drug over the study period. A small number of practices treated only 1 or 2 patients with either treatment and the maximum number of individuals treated by a practice was $1911$.

\begin{longtable}{l|l|l}
\hline
Variable                        & DPP4i & SGLT2i \\ 
                                 & $n = 109608$ & $n = 77229$ \\ \hline
\endhead
HbA1c (mmol/mol)                &  73.2 (14.1)    &  77.2 (14.8)        \\
BMI (kg/$m^2$)                  &  31.9 (6.58)	  &  33.8 (6.82)        \\
eGFR (ml/min/1.73$m^2$)         &  87.7 (18.8)    &  94.8 (15.1)        \\
ALT  (U/L)                      &  31.7 (19.4)    &  34.5 (20.1)        \\
Age (years)                     &  63.0 (12.5)    &  58.3 (10.5)        \\
T2D duration (years)            &  9.01 (6.54)    &  9.84 (6.34)        \\
Sex                             &                 &                      \\
\multicolumn{1}{r|}{female}     &  43973 (40.1$\%$) &   30022 (38.9$\%$) \\
\multicolumn{1}{r|}{male}     &  65635  (59.9$\%$) &   47207 (61.1$\%$) \\
Year of treatment prescription   &       &        \\
\multicolumn{1}{r|}{2013}       &  12651 (11.54$\%$)  &  1254 (1.62$\%$)  \\  
\multicolumn{1}{r|}{2014}       &  13038 (11.9$\%$)   &  5537 (7.17$\%$)  \\ 
\multicolumn{1}{r|}{2015}       &  14832 (13.53$\%$)  &  10155 (13.15$\%$) \\ 
\multicolumn{1}{r|}{2016}       &  16704 (15.24$\%$)  &  11121 (14.4$\%$)  \\ 
\multicolumn{1}{r|}{2017}       &  16960 (15.47$\%$)  &  12398 (16.05$\%$)  \\ 
\multicolumn{1}{r|}{2018}       &  16326 (14.89$\%$)  &  14325 (18.55$\%$)   \\ 
\multicolumn{1}{r|}{2019}       &  13969 (12.74$\%$)  &  15862 (20.54$\%$)   \\ 
\multicolumn{1}{r|}{2020}       &  5128 (4.68$\%$)    &  6577  (8.52$\%$)  \\ 
\newpage
Ethnicity                       &       &        \\
\multicolumn{1}{r|}{White}          &  84068 (76.7$\%$)   &  59393 (76.9$\%$)      \\
\multicolumn{1}{r|}{South Asian}    &  15026 (13.7$\%$)   &  10963 (14.2$\%$)      \\
\multicolumn{1}{r|}{Black}          &  5922 (5.4$\%$)     &  3352 (4.3$\%$)      \\
\multicolumn{1}{r|}{Other}          &  1647 (1.5$\%$)     &  1148 (1.5$\%$)      \\
\multicolumn{1}{r|}{Mixed}          &  1063 (1.0$\%$)     &  775 (1.0$\%$)     \\
Deprivation                     &    5.91 (2.85)   &  5.89 (2.86)      \\
Smoking status                  &       &        \\
\multicolumn{1}{r|}{Active smoker}  &   17771 (16.2$\%$)    &  12744 (16.5$\%$)      \\
\multicolumn{1}{r|}{Ex-smoker}      &   58610 (53.5$\%$)    &  41860 (54.2$\%$)      \\
\multicolumn{1}{r|}{Non-smoker}     &   27859 (25.4$\%$)   &   19451 (25.2$\%$)     \\
Number of concurrent treatments &       &        \\
\multicolumn{1}{r|}{1\textcolor{white}{+}}          &   11253 (10.3$\%$)    &   4293 (5.6$\%$)     \\
\multicolumn{1}{r|}{2\textcolor{white}{+}}          &   63519 (58.0$\%$)    &   33783 (43.7$\%$)     \\
\multicolumn{1}{r|}{3+}          &  34836 (31.8$\%$)     &   39153 (50.7$\%$)     \\
Line of treatment      &       &        \\
\multicolumn{1}{r|}{1\textcolor{white}{+}}     &  2461 (2.2$\%$)     &    571 (0.7$\%$)      \\
\multicolumn{1}{r|}{2\textcolor{white}{+}}          &  43887 (40.0$\%$)     &  14628 (18.9$\%$)      \\
\multicolumn{1}{r|}{3\textcolor{white}{+}}          &  44498 (40.6$\%$)     &  21041 (27.2$\%$)      \\
\multicolumn{1}{r|}{4+}         &  18762 (17.1$\%$)     &  40989 (53.1$\%$)	      \\
Patients ever taken Insulin      &       &        \\
\multicolumn{1}{r|}{yes}      &   5312 (4.85$\%$)    &  10691 (13.84$\%$)      \\
\caption{\label{tab:cohort_description} {Baseline characteristics of the CPRD T2D cohort for patients starting DPP4i (n = 109608) or SGLT2i (n = 77229) after 2013. Values are shown in mean (standard deviation) unless otherwise stated. Abbreviations: HbA1c (glycated haemoglobin), BMI (body mass index), eGFR (estimated glomerular rate), measured using the CKD-EPI Creatinine equation (2021), ALT (alanine aminotransferase), T2D (type 2 diabetes). Furthermore, deprivation was measured using the English Index of Multiple Deprivation (IMD) decile (1=most deprived, 10=least deprived).}}
\end{longtable}

\noindent Further data preparation of the study population was needed in order to apply all construction methods of $Z$ and the Two-Stage Least Squares IV estimation approach. For all construction methods a complete case dataset without missingness on the outcome variable was required. Additionally, for all construction methods other than the Ertefaie method and its extension method, a complete case dataset on the measured confounders was essential. In Table \ref{tab:missing_data}  in Appendix 4 an overview of the structure of missing values in the study population is given. Each construction method requires a different minimum number of patients treated by each provider ($n_{j,min}$). Providers with too little data are excluded from the analysis. A summary on $n_{j,min}$ together with information about the dataset sizes after exclusion of too small providers is given in Appendix 4 in Table \ref{tab:data_infromation}.

\subsection{Results}

\begin{figure}[H]
	\centering
	\includegraphics[width=0.7\linewidth]{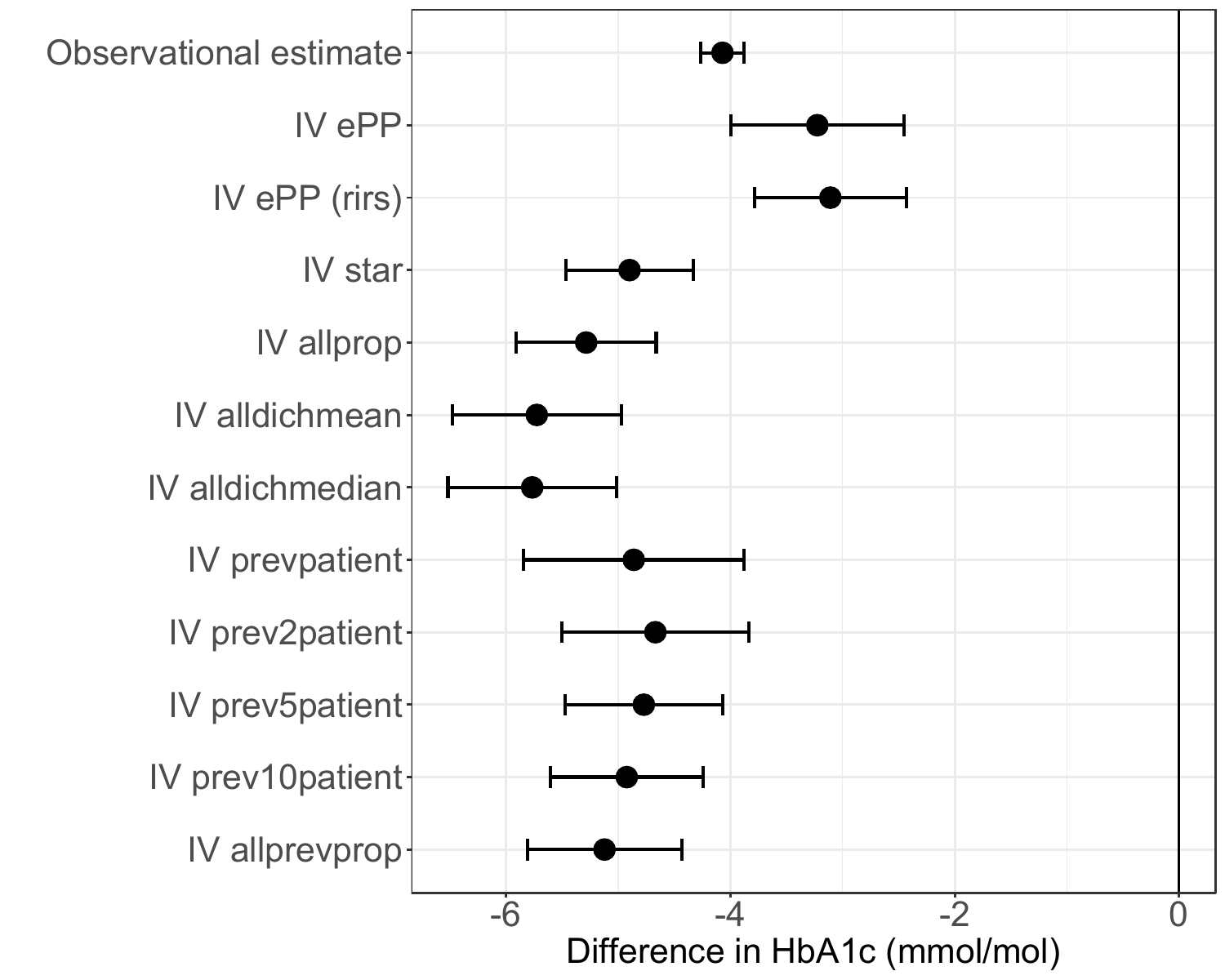}
	\caption{Estimation results of the relative treatment effect of SGLT2i versus DPP4i on the reduction of HbA1c (mmol/mol). Values smaller than 0 indicate that SGLT2i has a stronger HbA1c decreasing effect compared to DPP4i. Results are shown for a multivariable regression analysis (\textit{observational estimate}) and all IV estimates employing the construction methods of a preference-based IV.}
	\label{fig:results_application}
\end{figure}

\noindent It was possible to apply  all construction methods for the instrument explained in Section \ref{sec:ppIV_constructionmethods}  in this application  case study. Slightly different subsets of the study population were needed for each construction method due to different requirements on complete case data and minimum number of patients treated by each provider ($n_{j,min}$). From Table \ref{tab:SimAandB_additionalinfomation} it is clear that IV prev10patient used the smallest dataset for the analysis as this method requires a complete case dataset on the outcome variable and measured confounders and requires $n_{j,min} > 11$. Additionally, the data of the first 10 patients treated by each practice were excluded from the IV estimation as $Z$ cannot be calculated for these patients. \\
\\
The estimation results for the application case study are given in Figure \ref{fig:results_application} together with 95\% confidence intervals (CIs). CIs are taken from the outcome model, this ignores the uncertainty in the first stage model. All approaches show a consistently greater HbA1c reduction with SGLT2i compared to DPP4i. Compared with the other approaches, this difference is attenuated using the Ertefaie method and our extension method. From the simulation outlined in Section \ref{sec:simulation_study}, we have seen that we can trust these methods to account for non-ignorable missingness. As in the simulation study we see that our proposed method estimates the treatment effect with slightly more efficiency compared to IV ePP. All other methods shown in this plot rely on complete case analysis and therefore use smaller and potentially more selective datasets regarding patient characteristics. Using complete case analysis can lead to bias if the missingness is MNAR. \cite{Little02, Yang14, Ertefaie17}  Hence, it is possible that by using complete case analysis we exclude patients with certain characteristics and therefore overestimate the relative treatment effect. In Appendix 5 the estimation results for the case study are given for which all IV construction methods are applied to the same complete case dataset with $n_{j,min} > 11$. The results show that IV ePP and IV ePP (rirs) still lead to significantly smaller relative blood glucose benefit estimate compared to all of IV construction methods. Additionally, the Ertefaie method and its extension method utilize slightly different outcome models to estimate the treatment effect (second stage model of the IV estimation) because both methods only include confounders which are measured for all patients ($\mathbf{W_{obs}}$, as explained in Section \ref{sec:ppIV_Ertefaie} and Section \ref{sec:ppIV_Ertefaie_rirs}). As the estimation results for IV ePP and IV ePP (rirs) are in agreement, this might indicate that changing preferences of providers is less of an issue of concern in this analysis or that provider sizes are large enough for IV ePP to on average reflect suitably on $PP$.

\section{Discussion}\label{sec:discussion}

\begin{figure}[H]
	\hspace{-0.3cm}\includegraphics[width=1.04\linewidth, left]{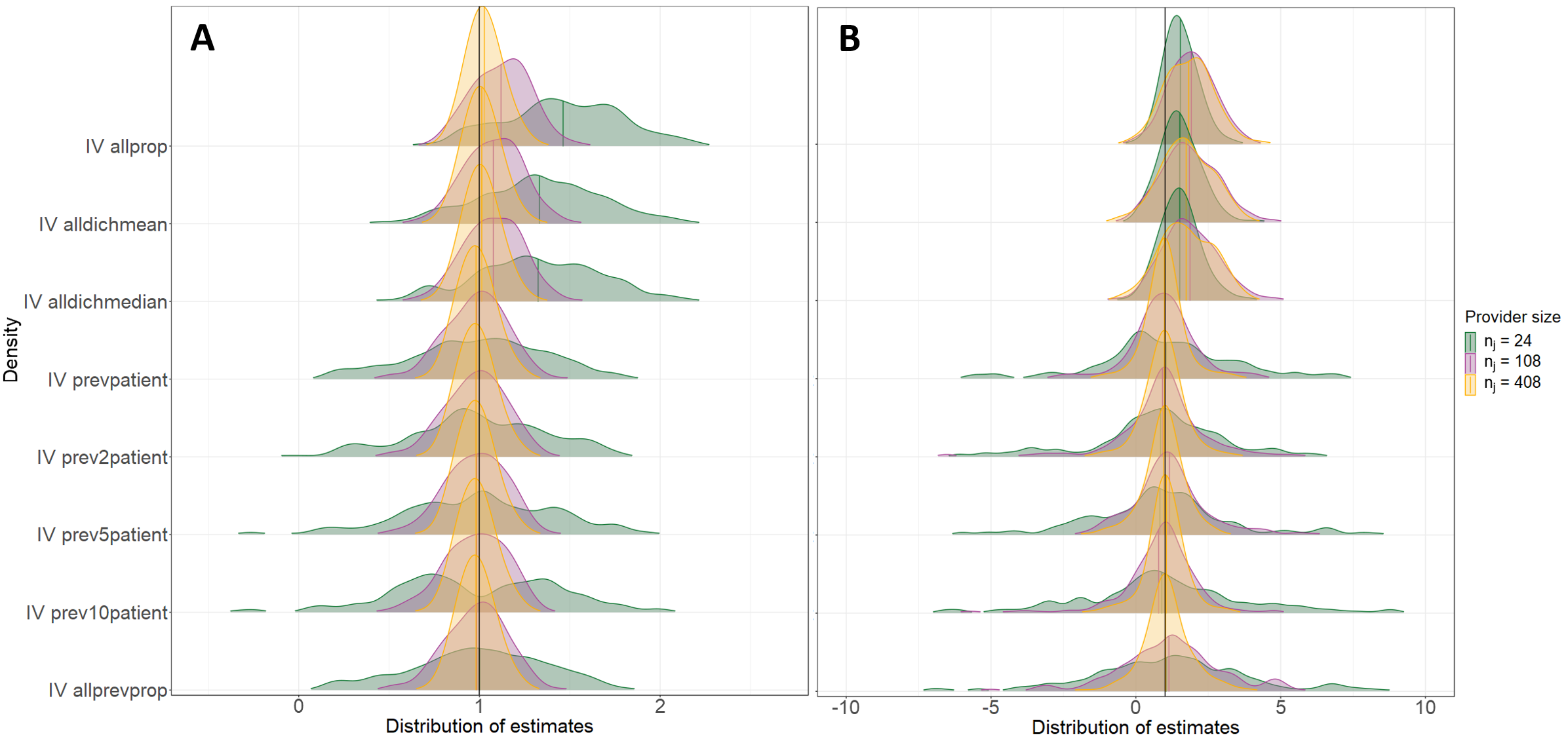}
	\caption{Estimation results of scenario 1: Change in provider size. Panel A: Estimation results for the data generation process using the Abrahamowicz model. Panel B: Estimation results for the data generation process using the extended Ertefaie model to simulate change in preference. Results are summarized for the rule-based construction methods of $Z$.}
	\label{fig:simulation_results_scenario1_rulebasedmethods}
\end{figure}

In this paper we conducted a state of the art performance analysis of the known construction methods for a preference-based instrument. With this study we add to the already existing literature on the performance evaluation of preference-based IVs  \cite{Abrahamowicz11, Ertefaie17, Uddin16, Ionescu09} by giving a comprehensive overview over all construction methods and evaluating all methods with respect to three important aspects: availability of prescription data within a provider, different missing data mechanisms for missing data in measured confounders and change in provider preference over time.  Additionally, we proposed an extended version of the construction method by Ertefaie et al. \cite{Ertefaie17} which aimed to combined the ability to deal with non-ignorable missingness and change in $PP$ using a random intercept and random slope model for the construction of $Z$. A simulation study was conducted using two different data generation strategies, to evaluate the performance independent of a specific generation process for $X$ that might benefit certain methods. Furthermore, all construction methods were showcased with a real life primary care dataset in a relative effectiveness study of two T2D oral agents. This case study outlines which data requirements are needed for each method to be applied to real life data, such as the necessity for complete case analysis, the exclusion of data due to insufficient prescription data or the inability to calculate a value for the surrogate instrument $Z$ with a specific construction method. \\
\\
Our results indicate that most model-based and rule-based construction methods do not have substantial problems in accounting for a change in provider preference. An exception from this are the rule-based methods which utilized all prescription information within a provider (IV allprop and its variations) and cannot reflect on change in preference appropriately. Our extension method models preference change with a random intercept random slope model. Results of the simulation study indicate that the method is able to estimate the treatment effect without bias. Additionally, in the simulation study and application case study the extension method estimated the treatment effect more efficiently compared to its original version proposed by Ertefaie et al. \cite{Ertefaie17}. Especially the performance of the more complex model-based approaches depend on the availability of sufficient data per provider. All mode-based methods struggled to estimate the treatment effect without bias in case of small provider sizes. When applying a model-based method to real life data the available data within providers is therefore a crucial consideration. More simple rule-based methods such as IV prevpatient proposed by Brookhart et al. \cite{Brookhart06b} could be considered as alternative in case of small datasets.  The Ertefaie method and our extension method are capable of estimating the true  treatment effect even in case of non-ignorable missingness in measured confounders. This makes them favorable in many observational research studies. Both construction methods will still require the use of complete case datasets based on the outcome variable, which may also lead to a selection of patients with specific characteristics and therefore a distortion of the treatment effect. All in all, the application case study showed the usefulness of triangulating results from different construction methods as proposed by Uddin et al. \cite{Uddin16}.  In doing so non-consistent results between the IV estimates can be discussed and sources of bias in the respective study may be discovered. \cite{Lawlor16} 
\\
\\
This study has some limitations which opens possibilities for further research. We have only tested the Ertefaie method and our extension on one specific mechanism to generate MNAR. With this, missingness was created based on the outcome variable value, the missing value itself, an unmeasured confounder and a provider level influence on missingness. We used the same generation process for missingness as in the original paper \cite{Ertefaie17}. It would be valuable to test both methods on different selection models for MNAR missingness to verify our findings and to explore sensitivity to miss-specification of the missingness model.  Results of the  simulation for the rule-based methods are presented in Appendix  2 and 3. When applying the construction methods to datasets with different provider sizes, the F-statistic results show that IV prevpatient and its variations are weak instruments with F-statistic values smaller than 10. Acceptable IV strength was only achieved by considering 5 or more previous prescription in the IV construction and for large provider sizes of $n_j = 408$. Interestingly, the estimation results for these construction methods did not show weak instrument bias (Figure \ref{fig:simulation_results_scenario1_rulebasedmethods}) as we would have expected. Further investigations will help to understand if the F-statistic results might be misleading for  this instrument, maybe due an introduction of serial autocorrelation between $Z$ and $X$ when using previous prescriptions to reflect on provider preference.  For the application case study we faced some data limitation using CPRD primary care data on the information available to construct a surrogate for provider preference. Only information on the allocation of patients to practices is given in the data. In reality, patients will be treated by different physicians within a practice with different prescription preference.  By constructing a surrogate on practice level, information of prescription pattern will be aggregated which will lead to measurement bias unless all physicians within a practice have the same preference.  
\\
\\
In summary, our study shows that IV methods using provider preference can be a useful tool for causal inference from observational health data, with both model-based and rule-based construction methods of preference-based instruments performing well in our simulation study as long as changes in provider preference over time are incorporated. Both the Ertefaie method and our proposed extension method are capable of estimating causal treatment effects even in case of non-ignorable missingness in measured confounders, and are recommended where sufficient data are available.

\subsection{Acknowledgements} 
This article is based in part on data from the Clinical Practice Research Datalink obtained under licence from the UK Medicines and Healthcare products Regulatory Agency. CPRD data is provided by patients and collected by the NHS as part of their care and support. This study was supported by the National Institute for Health and Care Research Exeter Biomedical Research Centre. The views expressed are those of the author(s) and not necessarily those of the NIHR or the Department of Health and Social Care.

\subsection{Funding}
Approval for the study was granted by the CPRD Independent Scientific Advisory Committee (ISCA 22\_002000). LMG, JB and JMD are supported by the UK Research and Innovation Expanding Excellence in England fund. BS and JMD receive funding from MRC and MASTERMIND. This research was funded by the Medical Research Council (UK) (MR/N00633X/1).

\subsection{Conflict of interest}
JB is a part time employee of Novo Nordisk. This project is unrelated to his work for the company.

\subsection{Author contribution}
LMG conceived the idea for the study and JB helped refining the research question of this manuscript. LMG and JB developed the proposed method for constructing preference-based instrumental variable. LMG conducted the analysis and drafted the original version of the paper which all authors helped to edit. All authors read and approved the final version of the manuscript.

\subsection{Data availability statement}
Data from CPRD is available to all researchers following successful application to the ISAC. Source code for this research for all simulations and the application study in this paper is available at \url{https://github.com/GuedemannLaura/ppIV}.
\newpage

\bibliographystyle{IEEEtranN}
\bibliography{references}

\newpage

\section*{Appendix 1}
Table \ref{tab:SimAandB_additionalinfomation} shows additional information on the data generation process for the simulation study explained in Section \ref{sec:simulation_study}. The left side of this table shows the variance of the simulated outcome variable $Y$ and the proportion of patients treated with $X_{ji} = 1$ (in $\%$) for the simulation strategy using the Abrahamowicz model to generate $X$. The right side of the table shows these results for the simulation which employs the extended Ertefaie model to simulate $X$. 

\begin{table}[H]
\begin{tabular}{ll|ll||ll}
\multicolumn{2}{l|}{\multirow{2}{*}{}}                         & \multicolumn{4}{c}{Simulation strategy for $X$ }  \\ \cline{3-6} 
\multicolumn{2}{l|}{\multirow{2}{*}{}}                         & \multicolumn{2}{l||}{Abrahamowicz model } & \multicolumn{2}{l}{Extended Ertefaie model} \\ \cline{3-6} 
\multicolumn{2}{l|}{}                                          & \multicolumn{1}{l|}{Var(Y)}    & $p(X_{ji} = 1)\times 100$    & \multicolumn{1}{l|}{Var(Y)}  & $p(X_{ji} = 1)\times100$  \\ \hline
\multicolumn{1}{c|}{\multirow{3}{*}{Scenario 1}} & $n_j = 24$  & \multicolumn{1}{l|}{7.798}          &               42.788          & \multicolumn{1}{l|}{6.996}        &        56.394                \\
\multicolumn{1}{c|}{}                            & $n_j = 108$ & \multicolumn{1}{l|}{7.778}          &           42.767              & \multicolumn{1}{l|}{6.992}        &        55.825                \\
\multicolumn{1}{c|}{}                            & $n_j = 408$ & \multicolumn{1}{l|}{7.782}          &       42.462                  & \multicolumn{1}{l|}{6.991}        &         56.085               \\ \hline
\multicolumn{1}{l|}{\multirow{3}{*}{Scenario 2}} & no NAs      & \multicolumn{1}{l|}{7.782}          &     42.462                    & \multicolumn{1}{l|}{6.991}        &        56.085                \\
\multicolumn{1}{l|}{}                            & MCAR        & \multicolumn{1}{l|}{7.783}          &     43.74                    & \multicolumn{1}{l|}{6.994}        &           56.302             \\
\multicolumn{1}{l|}{}                            & MNAR        & \multicolumn{1}{l|}{7.783}          &          43.732               & \multicolumn{1}{l|}{6.994}        &      56.301                  \\ \hline
\end{tabular}
\caption{\label{tab:SimAandB_additionalinfomation} Additional information on the outcome variance and the proportion of treated patients for both data generation strategies employed in the simulation explained in Section \ref{sec:simulation_study}.}
\end{table}


\section*{Appendix 2}

In Section \ref{sec:simulation_study} a state of the art simulation study is described which focuses on  three important aspects to consider when using provider preference-based IVs: The availability of prescription data for each provider of the study population, missing data in the measured confounders and possible change in provider preference over time. Results summary outlined in Section \ref{sec:simulation_study} focused on the model-based construction methods of $Z$. The additional results for all rule-based methods introduced in Section \ref{sec:ppIV_constructionmethods} are given below. Table \ref{tab:summary_methods_2} summarized the rule-based IV construction methods and their abbreviations. 

\begin{table}[H]
\begin{tabular}{l|p{12.5cm}}
\hline
Abbreviation     & Method  \\ \hline
IV allprop       & IV based on all prescriptions (proportion)                                         \\
IV alldichmean   & IV based on all prescriptions (dichotomized with mean)                             \\
IV alldichmedian & IV based on all prescriptions (dichotomized with median)                           \\
IV prevpatient   & IV based on previous prescription                                                  \\
IV prev2patient  & IV based on previous 2 prescriptions  \\
IV prev5patient  & IV based on previous 5 prescriptions  \\
IV prev10patient  & IV based on previous 10 prescriptions  \\
IV allprevprop   & IV based on all previous prescriptions  \\     \hline                             
\end{tabular}
\caption{\label{tab:summary_methods_2} Summary of rule-based construction methods applied in the simulation and their abbreviations.}
\end{table}

\noindent For scenario 1 and in Panel A, the construction methods which use a subset of previous prescriptions are unbiased for larger $n_j$ as they can reflect on a change in $PP$. We do not see and improvement of the estimation variance when including more previous prescription data in the construction of $Z$. But the F-statistic results in Table \ref{tab:f_stats_results_simulationA_rulebased} reflects that including more previous patient in the construction of $Z$ leads to stronger instruments. Similar results have been shown by  Uddin et al. \cite{Uddin16}. They concluded that IV prev10patient outperformed IV prev5patient and IV prevpatient with regards to strength. In contrast,  IV allprop and its variations cannot reflect on $PP$ because all prescription data within a provider is used simultaneously for the construction of $Z$. The estimation results are biased for $n_j = 24$ and $n_j = 108$. In Panel B, the results for the simulation strategy using the extended Ertefaie model to simulate $X$ are given. The results show generally larger estimation variance for methods using a subset of previous prescriptions and for smaller $n_j$. For larger $n_j$ the estimation variance look similar over all methods. As for the first simulation strategy, methods using all prescription information to construct $Z$ are biased even for larger $n_j$. This results is also reflected in their coverage rates summarized in Table \ref{tab:performance_measures_scenario1_simulationB}.

\begin{figure}[H]
	\hspace{-0.3cm}\includegraphics[width=1.04\linewidth, left]{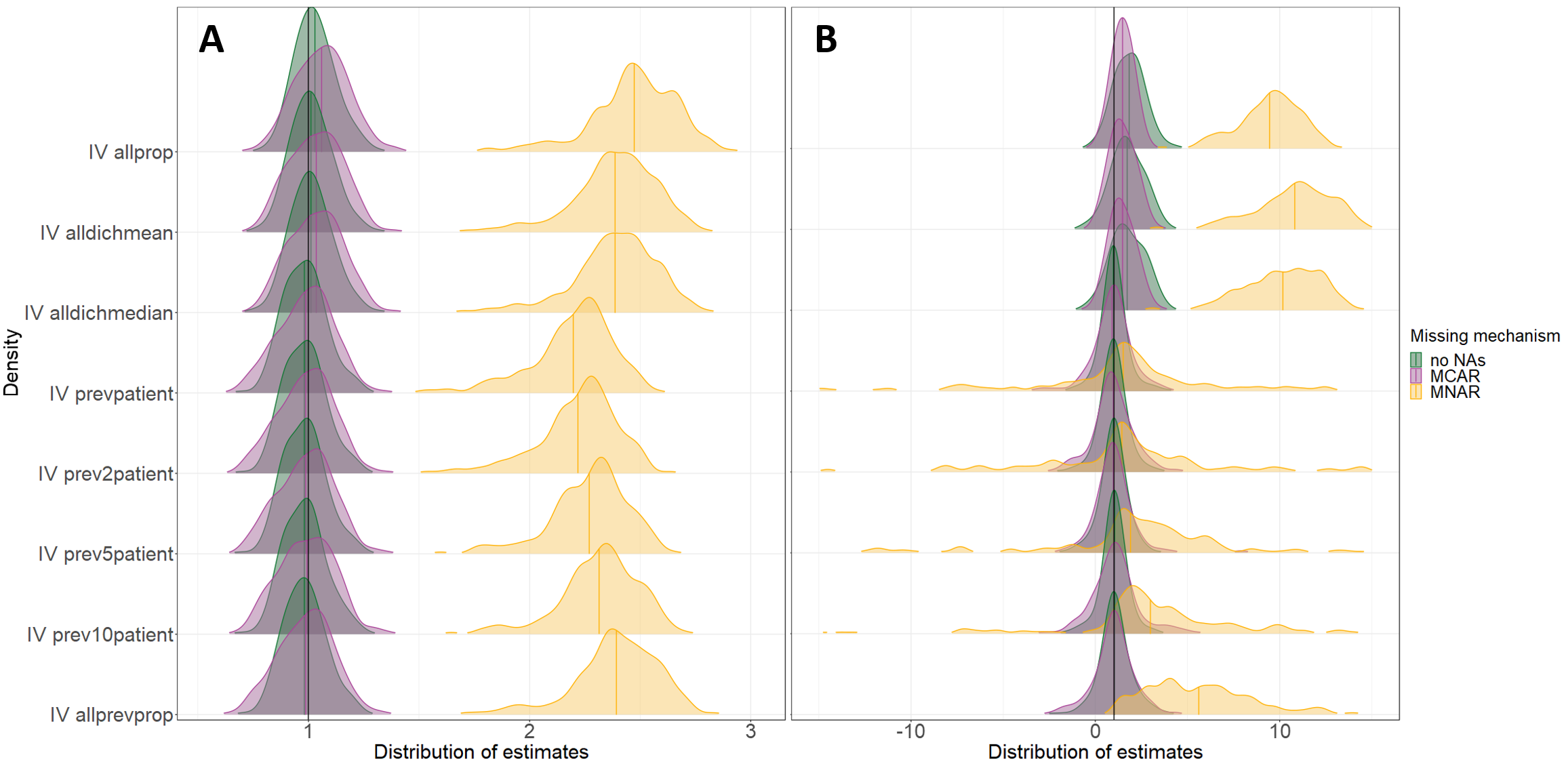}
	\caption{Estimation results of scenario 2: missing data mechanisms. Panel A: estimation results for the data generation process using the Abrahamowicz model. Panel B: estimation results for the data generation process using the extended Ertefaie model to simulate change in preference. Results are summarized for the rule-based construction methods of $Z$.}
	\label{fig:simulation_results_scenario2_rulebasedmethods}
\end{figure}

\noindent For scenario 2, all model-based construction methods result in biased treatment effect estimations in case of non-ignorable missingness. These results hold for both stimulation strategies.


\begin{table}[H]
	\begin{center}
		\begin{normalsize}
			\color{black}
			\begin{tabular}{r|rrc|rrc}
				\hline
				 &\multicolumn{3}{c|}{Scenario 1}&\multicolumn{3}{c}{Scenario 2}\\ 
				&\multicolumn{3}{c|}{Provider size ($n_j$)}&\multicolumn{3}{c}{Missing mechanism}\\ 
				&$24$&$108$&$408$&no NAs&MCAR&MNAR\\ 
	\hline
IV allprop & 63.35 & 95.81 & 234.39  & 234.39 & 160.32 & 93.49 \\ 
IV alldichmean & 42.42 & 62.81 & 158.18   & 158.18 & 107.13 & 48.24 \\ 
IV alldichmedian & 41.73 & 62.02 & 155.99  & 155.99 & 105.81 & 50.05 \\ 
IV prevpatient & 1.01 & 1.43 & 3.78   & 3.78  & 2.8 & 9.9 \\ 
IV prev2patient & 1.38  & 2.15 & 6.47   & 6.47 & 4.15 & 17.15 \\ 
IV prev5patient & 1.39 & 3.96 & 14.26  & 14.26 & 8.96 & 32.57 \\ 
IV prev10patient & 1.25 & 6.24 & 26.96   & 26.96  & 17.03 & 46.32 \\ 
IV allprevprop& 1.21 & 5.89 & 54.35   & 54.35 & 24.26  & 63.68 \\ 
\hline
			\end{tabular}
		\end{normalsize}
	\end{center}
\caption{\label{tab:f_stats_results_simulationA_rulebased} F-statistic results for the instrument $Z$ from the first stage regression model of the Two-Stage Least Squares approach. The results are summarized for all scenarios. For this simulation the treatment decision $X$ using the Abrahamowicz model. This table summarized the results of all rule-based construction methods for $Z$.}
\end{table}
\color{black}


\begin{table}[H]
	\begin{center}
		\begin{normalsize}
			\color{black}
			\begin{tabular}{r|rrc|rrc}
				\hline
				 &\multicolumn{3}{c|}{Scenario 1}&\multicolumn{3}{c}{Scenario 2}\\ 
				&\multicolumn{3}{c|}{Provider size ($n_j$)}&\multicolumn{3}{c}{Missing mechanism}\\ 
				&$24$&$108$&$408$&no NAs&MCAR&MNAR\\ 
	\hline
IV allprop & 75.26 & 29.05 & 26.43  & 26.43 & 52.03 & 93.93 \\ 
IV alldichmean & 51.7 & 18.79 & 17  & 17 & 33.91 & 62.63 \\ 
IV alldichmedian & 50.16 & 18.4 & 16.8  & 16.8 & 33.55  & 63.38 \\ 
IV prevpatient &  8.2 & 44.95 & 186.92  & 186.92 & 46.36 & 62.37 \\ 
IV prev2patient & 9.05 & 45.77 & 190.14   & 190.14 & 41.15 & 60.2 \\ 
IV prev5patient & 8.12 & 44.32 & 156.08  & 156.08 & 39.72 & 65.6 \\ 
IV prev10patient & 7.42 & 37.61 & 136.51   & 136.51 & 39.07 & 74.18 \\ 
IV allprevprop & 3.21 & 10.25 & 90.62   & 90.62 & 39.61 & 213.96 \\ 
\hline
			\end{tabular}
		\end{normalsize}
	\end{center}
\caption{\label{tab:f_stats_results_simulationB_rulebased} F-statistic results for the instrument $Z$ from the first stage regression model of the Two-Stage Least Squares approach. The results are summarized for all scenarios. For this simulation the treatment decision $X$ is generated using the extended Ertefaie model. This table summarized the results of all rule-based construction methods for $Z$.}
\end{table}
\color{black}


\newpage
\section*{Appendix 3}

In the following additional estimation performance measures for the simulation study are summarized. Results on the bias, standard error (SE) coverage (in \%) and root mean squared error (RMSE) are given for all scenarios and proxy measure construction methods.


\begin{longtable}{l|l|rrrr}
  \hline
Method & $n_j$ & Bias & SE & Coverage & RMSE \\ 
  \hline
  \endhead
\multirow{3}{*}{Obs. estimate} & 24 & 1.5552 & 0.0083 & 0 & 1.5596 \\ 
 & 108 & 1.545 & 0.0041 & 0 & 1.5461 \\ 
 & 408 & 1.5427 & 0.0026 & 0 & 1.5431 \\ 
 \hline
\multirow{3}{*}{IV(PP)} & 24 & -0.0119 & 0.0236 & 98.5 & 0.3331 \\ 
 & 108 & -0.0173 & 0.0109 & 98 & 0.1548 \\ 
& 408 & -0.0149 & 0.0061 & 97.5 & 0.0867 \\ 
\hline
\multirow{3}{*}{IV(PP) cc} & 24 & -0.0119 & 0.0236 & 98.5 & 0.3331 \\ 
& 108 & -0.0173 & 0.0109 & 98 & 0.1548 \\ 
& 408 & -0.0149 & 0.0061 & 97.5 & 0.0867 \\ 
\hline
\multirow{3}{*}{IV ePP} & 24 & 0.4321 & 0.0258 & 74 & 0.5649 \\ 
& 108 & 0.1196 & 0.011 & 96.5 & 0.1957 \\ 
& 408 & 0.0207 & 0.0065 & 95.5 & 0.0942 \\ 
\hline
\multirow{3}{*}{IV ePP (rirs)} & 24 & 0.3987 & 0.0285 & 75.5 & 0.5664 \\ 
& 108 & 0.1681 & 0.0111 & 84 & 0.2298 \\ 
& 408 & 0.0355 & 0.0061 & 94.5 & 0.0933 \\ 
\hline
\multirow{3}{*}{IV star} & 24 & 0.2462 & 0.0248 & 93 & 0.4281 \\ 
& 108 & 0.1111 & 0.0119 & 93.5 & 0.2007 \\ 
& 408 & 0.0169 & 0.0064 & 96 & 0.0918 \\ 
\hline
\multirow{3}{*}{IV allprop} & 24 & 0.4641 & 0.0228 & 71 & 0.5649 \\ 
 & 108 & 0.12 & 0.0111 & 95.5 & 0.1968 \\ 
 & 408 & 0.0287 & 0.0064 & 94.5 & 0.0948 \\ 
 \hline
\multirow{3}{*}{IV alldichmean} & 24 & 0.3329 & 0.0241 & 87 & 0.476 \\ 
 & 108 & 0.0766 & 0.0112 & 96.5 & 0.1759 \\ 
 & 408 & 0.0119 & 0.0066 & 95 & 0.0932 \\ 
 \hline
\multirow{3}{*}{IV alldichmedian} & 24 & 0.3259 & 0.0243 & 87.5 & 0.473 \\ 
 & 108 & 0.0758 & 0.0113 & 96.5 & 0.1759 \\ 
 & 408 & 0.0119 & 0.0066 & 95 & 0.0938 \\ 
 \hline
\multirow{3}{*}{IV prevpatient} & & 0.0022 & 0.0265 & 98.5 & 0.3735 \\ 
 & 108 & -0.0166 & 0.0119 & 97.5 & 0.1691 \\ 
 & 408 & -0.0179 & 0.0067 & 97.5 & 0.0964 \\ 
 \hline
\multirow{3}{*}{IV prev2patient} & 24 & -0.0014 & 0.0272 & 98 & 0.3835 \\ 
 & 108 & -0.0178 & 0.012 & 97.5 & 0.1697 \\ 
 & 408 & -0.0182 & 0.0067 & 98 & 0.0959 \\
 \hline
\multirow{3}{*}{IV prev5patient} & 24 & 0.0001 & 0.0302 & 97 & 0.4256 \\ 
 & 108 & -0.0166 & 0.0120 & 98 & 0.1699 \\ 
 & 408 & -0.0185 & 0.0067 & 98 & 0.0957 \\ 
 \hline
 \pagebreak
\multirow{3}{*}{IV prev10patient} & 24 & -0.002 & 0.0334 & 97 & 0.4716 \\ 
 & 108 & -0.0171 & 0.0123 & 98 & 0.1737 \\ 
 & 408 & -0.0175 & 0.0067 & 97 & 0.0965 \\ 
 \hline
\multirow{3}{*}{IV allprevprop} & 24 & 0.0012 & 0.0263 & 98 & 0.3717 \\ 
& 108 & -0.0154 & 0.0117 & 97.5 & 0.1664 \\ 
 & 408 & -0.0173 & 0.0066 & 97.5 & 0.0949 \\
\caption{\label{tab:performance_measures_scenario1_simulationA} Summary of performance measures for scenario 1. For this simulation the treatment decision $X$ is generated using the Abrahamowicz model.}
\end{longtable}


\begin{longtable}{l|l|rrrr}
  \hline
Method & Missing mechanism & Bias & SE & Coverage & RMSE \\ 
  \hline
  \endhead
\multirow{3}{*}{obs. estimate} & no NAs & 1.5427 & 0.0026 & 0 & 1.5431 \\ 
 & MCAR & 1.5508 & 0.0029 & 0 & 1.5513 \\ 
 & MNAR & 1.6638 & 0.0032 & 0 & 1.6644 \\ 
 \hline
\multirow{3}{*}{IV(PP)} & no NAs & -0.0149 & 0.0061 & 97.5 & 0.0867 \\ 
 & MCAR & -0.0143 & 0.0063 & 96 & 0.0893 \\ 
 & MNAR & -0.0143 & 0.0063 & 0 & 0.0893 \\ 
 \hline
\multirow{3}{*}{IV(PP) cc} & no NAs & -0.0149 & 0.0061 & 97.5 & 0.0867 \\ 
 & MCAR & -0.0109 & 0.0082 & 96 & 0.1157 \\ 
& MNAR & 1.0428 & 0.013 & 96 & 1.0588 \\ 
\hline
\multirow{3}{*}{IV ePP} & no NAs & 0.0207 & 0.0065 & 95.5 & 0.0942 \\ 
 & MCAR & 0.0712 & 0.0182 & 94.5 & 0.2667 \\ 
 & MNAR & 0.0402 & 0.0191 & 95 & 0.2729 \\ 
 \hline
\multirow{3}{*}{IV ePP (rirs)} & no NAs & 0.0355 & 0.0061 & 94.5 & 0.0933 \\ 
 & MCAR & 0.1206 & 0.0157 & 95.5 & 0.2525 \\ 
& MNAR & 0.1063 & 0.0167 & 95 & 0.2586 \\ 
\hline
\multirow{3}{*}{IV star} & no NAs & 0.0169 & 0.0064 & 96 & 0.0918 \\ 
 & MCAR & 0.0518 & 0.009 & 95 & 0.1373 \\ 
 & MNAR & 1.4479 & 0.0137 & 0 & 1.4607 \\
 \hline
\multirow{3}{*}{IV allprop} & no NAs & 0.0287 & 0.0064 & 94.5 & 0.0948 \\ 
 & MCAR & 0.0601 & 0.0085 & 93 & 0.1342 \\ 
& MNAR & 1.4727 & 0.0137 & 0 & 1.4853 \\ 
\hline
\multirow{3}{*}{IV alldichmean} & no NAs & 0.0119 & 0.0066 & 95 & 0.0932 \\ 
 & MCAR & 0.0349 & 0.0085 & 97 & 0.1253 \\ 
 & MNAR & 1.386 & 0.0134 & 0 & 1.3988 \\ 
 \hline
\multirow{3}{*}{IV alldichmedian} & no NAs & 0.0119 & 0.0066 & 95 & 0.0938 \\ 
 & MCAR & 0.0339 & 0.0085 & 97 & 0.1248 \\ 
 & MNAR & 1.3856 & 0.0134 & 0 & 1.3985 \\ 
 \hline
\multirow{3}{*}{IV prevpatient} & no NAs & -0.0179 & 0.0067 & 97.5 & 0.0964 \\ 
& MCAR & -0.0117 & 0.0089 & 95 & 0.1256 \\ 
 & MNAR & 1.1966 & 0.0132 & 0 & 1.2109 \\ 
 \hline
  \pagebreak
 \multirow{3}{*}{IV prev2patient} & no NAs & -0.0182 & 0.0067 & 98 & 0.0959 \\ 
 & MCAR & -0.0112 & 0.0089 & 95.5 & 0.1257 \\ 
 & MNAR & 1.2198 & 0.013 & 0 & 1.2335 \\ 
 \hline
\multirow{3}{*}{IV prev5patient} & no NAs & -0.0185 & 0.0067 & 98 & 0.0957 \\ 
 & MCAR & -0.0112 & 0.0089 & 95 & 0.1261 \\ 
 & MNAR & 1.2689 & 0.0126 & 0 & 1.2813 \\ 
 \hline
\multirow{3}{*}{IV prev10patient} & no NAs & -0.0175 & 0.0067 & 97 & 0.0965 \\ 
 & MCAR & -0.0078 & 0.009 & 96.5 & 0.1265 \\ 
 & MNAR & 1.3157 & 0.0128 & 0 & 1.328 \\ 
 \hline
\multirow{3}{*}{IV allprevprop} & no NAs & -0.0173 & 0.0066 & 97.5 & 0.0949 \\ 
 & MCAR & -0.0112 & 0.0089 & 95.5 & 0.1258 \\ 
  & MNAR & 1.393 & 0.0134 & 0 & 1.4058 \\ 
\caption{\label{tab:performance_measures_scenario2_simulationA} Summary of performance measures for scenario 2. For this simulation the treatment decision $X$ is generated using the Abrahamowicz model.}
\end{longtable}


\begin{longtable}{l|l|rrrr}
  \hline
Method & $n_j$ & Bias & SE & Coverage & RMSE \\ 
  \hline
  \endhead
\multirow{3}{*}{obs. estimate} & 24 & 0.3568 & 0.0063 & 3.5 & 0.3677 \\ 
  & 108 & 0.3664 & 0.0039 & 0 & 0.3705 \\ 
 & 408 & 0.3616 & 0.0027 & 0 & 0.3637 \\ 
 \hline
\multirow{3}{*}{IV(PP)} & 24 & -0.0191 & 0.008 & 96 & 0.1138 \\ 
  & 108 & -0.0039 & 0.0039 & 94.5 & 0.0549 \\ 
   & 408 & -0.009 & 0.002 & 93.5 & 0.0303 \\ 
   \hline
\multirow{3}{*}{IV(PP) cc} & 24 & -0.0191 & 0.008 & 96 & 0.1138 \\ 
  & 108 & -0.0039 & 0.0039 & 94.5 & 0.0549 \\ 
  & 408 & -0.009 & 0.002 & 93.5 & 0.0303 \\ 
  \hline
\multirow{3}{*}{IV ePP} & 24 & 0.2263 & 0.3708 & 92.5 & 5.235 \\ 
  & 108 & 0.293 & 0.2243 & 93 & 3.1779 \\ 
  & 408 & 0.1837 & 0.1109 & 90.5 & 1.5754 \\ 
  \hline
\multirow{3}{*}{IV ePP (rirs)} & 24 & 0.1751 & 0.1099 & 97 & 1.5605 \\ 
  & 108 & 0.1957 & 0.0726 & 94 & 1.0433 \\ 
  & 408 & 0.0105 & 0.044 & 96.5 & 0.6201 \\
  \hline
\multirow{3}{*}{IV star} & 24 & 0.3880 & 0.0538 & 95 & 0.8521 \\ 
  & 108 & 0.3638 & 0.0553 & 88.5 & 0.8609 \\ 
  & 408 & 0.3506 & 0.0442 & 86 & 0.7155 \\
  \hline
\multirow{3}{*}{IV allprop} & 24 & 0.5342 & 0.0394 & 82 & 0.7713 \\ 
  & 108 & 0.9229 & 0.0547 & 78 & 1.2028 \\ 
  & 408 & 0.8371 & 0.0582 & 78.5 & 1.1725 \\ 
  \hline
\multirow{3}{*}{IV alldichmean} & 24 & 0.524 & 0.049 & 89.5 & 0.8677 \\ 
  & 108 & 0.8589 & 0.0681 & 86.5 & 1.2882 \\ 
  & 408 & 0.7379 & 0.0634 & 85.5 & 1.1597 \\ 
  \hline
   \pagebreak
\multirow{3}{*}{IV alldichmedian} & 24 & 0.5228 & 0.0483 & 89.5 & 0.8583 \\ 
  & 108 & 0.874 & 0.0676 & 87 & 1.2934 \\ 
  & 408 & 0.7461 & 0.0652 & 84 & 1.1842 \\ 
  \hline
\multirow{3}{*}{IV prevpatient}& 24 & 0.0733 & 0.2228 & 95 & 3.1432 \\ 
  & 108 & 0.0188 & 0.0873 & 96 & 1.2324 \\ 
  & 408 & -0.0135 & 0.0529 & 96 & 0.7461 \\ 
  \hline
\multirow{3}{*}{IV prev2patient} & 24 & 0.071 & 0.2268 & 95.5 & 3.2 \\ 
  & 108 & -0.088 & 0.1119 & 95 & 1.5803 \\ 
 & 408 & -0.017 & 0.0595 & 94.5 & 0.8392 \\ 
 \hline
\multirow{3}{*}{IV prev5patient} & 24 & 0.1868 & 0.3076 & 96.5 & 4.3437 \\ 
  & 108 & 0.1625 & 0.0952 & 96.5 & 1.3533 \\ 
  & 408 & -0.0259 & 0.0513 & 92.5 & 0.7237 \\ 
  \hline
\multirow{3}{*}{IV prev10patient} & 24 & 0.2317 & 0.4096 & 96 & 5.7833 \\ 
  & 108 & -0.1494 & 0.1164 & 96.5 & 1.6494 \\ 
  & 408 & 0.0448 & 0.0558 & 93.5 & 0.7888 \\ 
  \hline
\multirow{3}{*}{IV allprevprop} & 24 & -0.0751 & 0.2498 & 98 & 3.525 \\ 
 & 108 & 0.2378 & 0.158 & 96.5 & 2.2412 \\ 
  & 408 & 0.0382 & 0.0677 & 94 & 0.9563 \\ 
\caption{\label{tab:performance_measures_scenario1_simulationB} Summary of performance measures for scenario 1. For this simulation the treatment decision $X$ is generated using the extended Ertefaie model.}
\end{longtable}


\begin{longtable}{l|l|rrrr}
  \hline
Method & Missing mechanism & Bias & SE & Coverage & RMSE \\ 
  \hline
  \endhead
\multirow{3}{*}{obs. estimate} & no NAs & 0.3568 & 0.0063 & 3.5 & 0.3677 \\ 
  & MCAR & 0.3664 & 0.0039 & 0 & 0.3705 \\ 
  & MNAR & 0.3616 & 0.0027 & 0 & 0.3637 \\ 
  \hline
\multirow{3}{*}{IV(PP)} & no NAs & -0.0191 & 0.008 & 96 & 0.1138 \\ 
  & MCAR & -0.0039 & 0.0039 & 94.5 & 0.0549 \\ 
  & MNAR & -0.009 & 0.002 & 93.5 & 0.0303 \\ 
  \hline
\multirow{3}{*}{IV(PP) cc} & no NAs & -0.0191 & 0.008 & 96 & 0.1138 \\ 
  & MCAR & -0.0039 & 0.0039 & 94.5 & 0.0549 \\ 
  & MNAR & -0.009 & 0.002 & 93.5 & 0.0303 \\ 
  \hline
\multirow{3}{*}{IV ePP} & no NAs & 0.2263 & 0.3708 & 92.5 & 5.235 \\ 
  & MCAR & 0.293 & 0.2243 & 93 & 3.1779 \\ 
  & MNAR & 0.1837 & 0.1109 & 90.5 & 1.5754 \\
  \hline
\multirow{3}{*}{IV ePP (rirs)} & no NAs & 0.1751 & 0.1099 & 97 & 1.5605 \\ 
  & MCAR & 0.1957 & 0.0726 & 94 & 1.0433 \\ 
  & MNAR & 0.0105 & 0.044 & 96.5 & 0.6201 \\ 
  \hline
\multirow{3}{*}{IV star} & no NAs & 0.388 & 0.0538 & 95 & 0.8521 \\ 
 & MCAR & 0.3638 & 0.0553 & 88.5 & 0.8609 \\ 
  & MNAR & 0.3506 & 0.0442 & 86 & 0.7155 \\ 
  \hline
   \pagebreak
\multirow{3}{*}{IV allprop} & no NAs & 0.5342 & 0.0394 & 82 & 0.7713 \\ 
  & MCAR & 0.9229 & 0.0547 & 78 & 1.2028 \\ 
  & MNAR & 0.8371 & 0.0582 & 78.5 & 1.1725 \\ 
  \hline
\multirow{3}{*}{IV alldichmean} & no NAs & 0.524 & 0.049 & 89.5 & 0.8677 \\ 
  & MCAR & 0.8589 & 0.0681 & 86.5 & 1.2882 \\ 
  & MNAR & 0.7379 & 0.0634 & 85.5 & 1.1597 \\ 
  \hline
\multirow{3}{*}{IV alldichmedian} & no NAs & 0.5228 & 0.0483 & 89.5 & 0.8583 \\ 
 & MCAR & 0.874 & 0.0676 & 87 & 1.2934 \\ 
  & MNAR & 0.7461 & 0.0652 & 84 & 1.1842 \\ 
  \hline
\multirow{3}{*}{IV prevpatient} & no NAs & 0.0733 & 0.2228 & 95 & 3.1432 \\ 
  & MCAR & 0.0188 & 0.0873 & 96 & 1.2324 \\ 
  & MNAR & -0.0135 & 0.0529 & 96 & 0.7461 \\ 
  \hline
\multirow{3}{*}{IV prev2patient} & no NAs & 0.0710 & 0.2268 & 95.5 & 3.2 \\ 
  & MCAR & -0.088 & 0.1119 & 95. & 1.5803 \\ 
 & MNAR & -0.017 & 0.0595 & 94.5 & 0.8392 \\
 \hline
\multirow{3}{*}{IV prev5patient} & no NAs & 0.1868 & 0.3076 & 96.5 & 4.3437 \\ 
  & MCAR & 0.1625 & 0.0952 & 96.5 & 1.3533 \\ 
  & MNAR & -0.0259 & 0.0513 & 92.5 & 0.7237 \\ 
  \hline
\multirow{3}{*}{IV prev10patient} & no NAs & 0.2317 & 0.4096 & 96 & 5.7833 \\ 
  & MCAR & -0.1494 & 0.1164 & 96.5 & 1.6494 \\ 
  & MNAR & 0.0448 & 0.0558 & 93.5 & 0.7888 \\ 
  \hline
\multirow{3}{*}{IV allprevprop} & no NAs & -0.0751 & 0.2498 & 98 & 3.525 \\ 
  & MCAR & 0.2378 & 0.158 & 96.5 & 2.2412 \\ 
  & MNAR & 0.0382 & 0.0677 & 94 & 0.9563 \\
\caption{\label{tab:performance_measures_scenario2_simulationB} Summary of performance measures for scenario 2. For this simulation the treatment decision $X$ is generated using the extended Ertefaie model.}
\end{longtable}


\newpage
\section*{Appendix 4}
A summary of the missing data for the application study explained in Section \ref{sec:application} is given in Table \ref{tab:missing_data} for the outcome variable and all measured confounders with missing values. For most of the Two-Stage Least Squares estimation, complete case datasets are necessary, except for the Ertefaie method and its extension method. Additionally, practices with too little data $n_j$ for the respective $Z$ construction method are excluded from the analysis. Based on the original study population and for each construction method, a separate dataset is therefore prepared to apply to the IV estimation procedure. 

\begin{longtable}{l|l|l|l}
\hline
Variable                        & DPP4i & SGLT2i & Overall\\  \hline
\endhead
Achieved HbA1c (mmol/mol)       &   34087 (31.01$\%$)  & 30189 (39.09$\%$)  & 64276 (34.4$\%$)\\
HbA1c (mmol/mol)                &  11559 (10.5$\%$) & 10379 (13.4$\%$) & 21938 (11.7$\%$) \\
BMI (kg/$m^2$)                  &  5689 (5.2$\%$)   & 3017 (3.9$\%$)  &  8706 (4.7$\%$) \\
eGFR (ml/min/1.73$m^2$)         &  726 (0.7$\%$)    &  347 (0.4$\%$)  & 1073 (0.6$\%$)  \\   
ALT (U/L)                       &  7513 (6.9$\%$)   &  4783 (6.2$\%$)  &  12296 (6.6$\%$)  \\    
Ethnicity                       &  1882 (1.7$\%$)   &    1598 (2.1$\%$) & 3480 (1.9$\%$)  \\ 
Deprivation                     &  63 (0.1$\%$)     &   45 (0.1$\%$) &  108 (0.1$\%$)    \\ 
Smoking status                  &  5368 (4.9$\%$)   &  3174 (4.1$\%$)   & 8542 (4.6$\%$)  \\ 
\caption{\label{tab:missing_data} {Summary of missing values in the study population. The summary shows the outcome variable (achieved HbA1cin mmol/mol) and all measured confounders with missing values.}}
\end{longtable}

\newpage
\begin{longtable}{l|l|l|l|l}
\hline
Method & Data size ($N$) & $J$ & $n_{j,min}$ & Complete case information\\  \hline
\endhead
 \multirow{1}{*}{Observational estimate}   & 103611   &  1403  &  -  & complete case on $Y$ and $W$\\
 & SGLT2i: 38.79 $\%$ & & &  \\ \hline
  IV ePP                 & 122556    &  1407  &  2  & complete case for $Y$\\
  & SGLT2i: 38.38 $\%$ & & &  \\ \hline
  IV ePP (rirs)          & 122556    &  1407  &  2  & complete case for $Y$ \\
  & SGLT2i: 38.38 $\%$ & & &  \\ \hline
  IV star                & 101377    &  1364  &  5  & complete case for $Y$ and $W$\\ 
  & SGLT2i: 38.95 $\%$ & & &  \\ \hline
  IV allprop             & 103598    &  1390  &  2  & complete case for $Y$ and $W$\\
  & SGLT2i: 38.79 $\%$ & & &  \\ \hline
  IV alldichmean         & 103598    &  1390  &  2  & complete case for $Y$ and $W$\\ 
  & SGLT2i: 38.79 $\%$ & & &  \\ \hline
  IV alldichmedian       & 103598    &  1390  &  2  & complete case for $Y$ and $W$\\ 
  & SGLT2i: 38.79 $\%$ & & &  \\ \hline
  IV prevpatient         & 102208    &  1390  &  2  & complete case for $Y$ and $W$\\
  & SGLT2i: 39.25 $\%$ & & &  \\ \hline
  IV prev2patient        & 100818    &  1373  &  3  & complete case for $Y$ and $W$\\
  & SGLT2i: 39.68 $\%$ & & &  \\ \hline
  IV prev5patient        & 96712     &  1358  &  6  & complete case for $Y$ and $W$\\
  & SGLT2i: 40.83 $\%$ & & &  \\ \hline
  IV prev10patient       & 89979     &  1326  & 11  & complete case for $Y$ and $W$\\ 
  & SGLT2i: 42.39 $\%$ & & &  \\ \hline
  IV allprevprop         & 102208    &  1390  &  2  & complete case for $Y$ and $W$\\ 
  & SGLT2i: 38.79 $\%$ & & &  \\ \hline
\caption{\label{tab:data_infromation} {Summary of the data preparation for the respective construction methods and the IV estimation. The data preparation process results in different study population sizes ($N$) and number of providers per dataset ($J$). Additionally, information on the minimum practice size ($n_{j,min}$) needed ton construct $Z$ and information on the complete case dataset construction is given.}}
\end{longtable}


\newpage
\section*{Appendix 5}
\begin{figure}[H]
	\centering
	\includegraphics[width=0.7\linewidth]{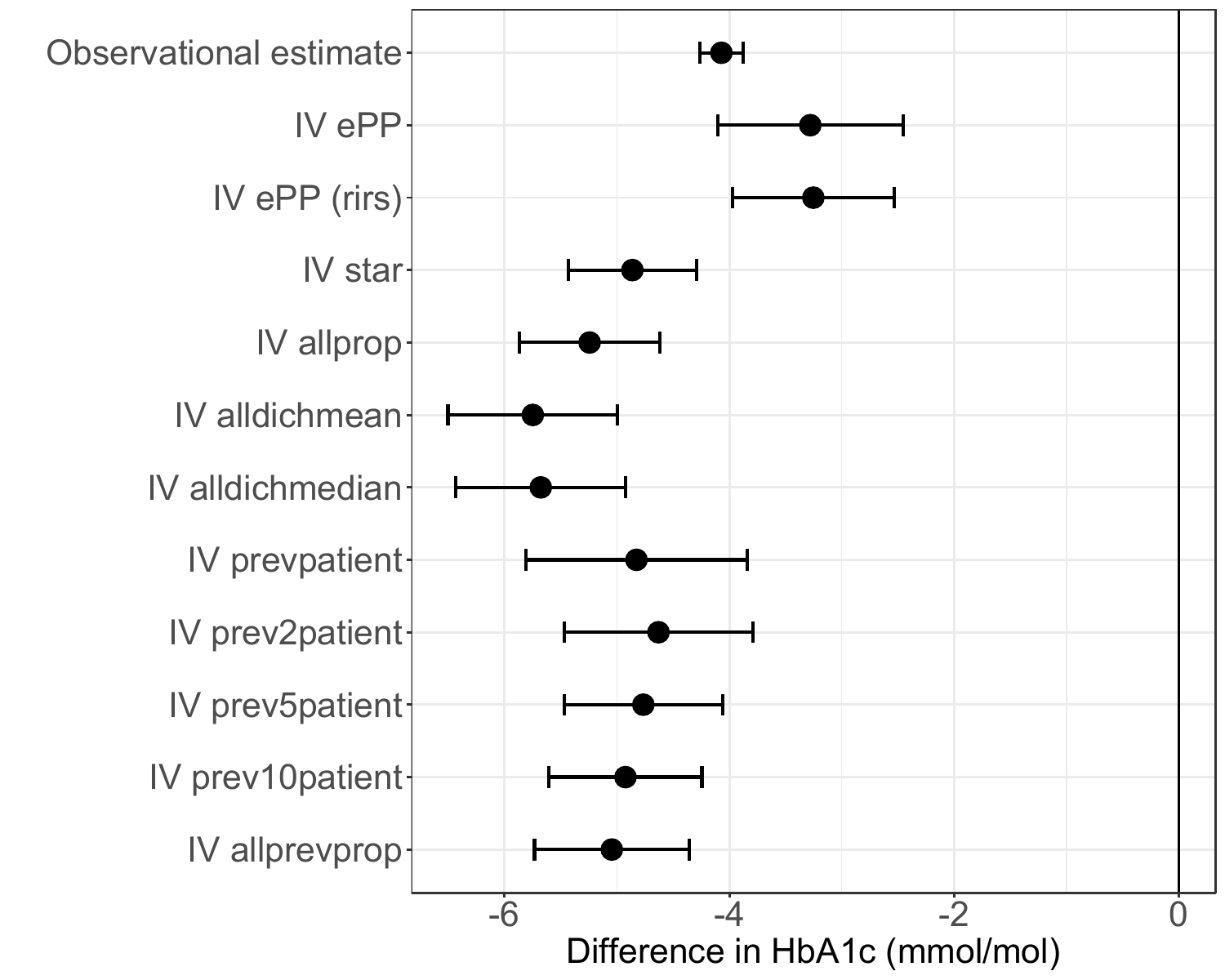}
	\caption{Estimation results of the relative treatment effect of SGLT2i versus DPP4i on the reduction of HbA1c (mmol/mol). Values smaller than 0 indicate that SGLT2i has a stronger HbA1c decreasing effect compared to DPP4i. Results are shown for a multivariable regression analysis (\textit{observational estimate}) and all IV estimates employing the construction methods of a preference-based IV. Estimation procedures was applied on the same complete case dataset with sufficient treatment prescription data for all IV construction methods.}
	\label{fig:results_application_completecase}
\end{figure}

The estimation results of the application case study outlines in Section \ref{sec:application} are represented in Figure \ref{fig:results_application}. These results are estimated using specific datasets for each of the construction method for $Z$ that depend on the complete case data and minimum practice size $n_{j,min}$ required for each of the methods. Only IV ePP and IV ePP (rirs) do not require a complete case dataset on the measured confounders. The two methods show significantly different estimation results compared to other IV methods. The analysis was therefore repeated on a complete case dataset with $n_{j,min} > 11$ on which all IV construction methods can be applied. Figure \ref{fig:results_application_completecase} summarizes the results of this analysis. It is noticeable that IV ePP and IV ePP (rirs) still lead to significantly smaller relative blood glucose benefit estimates compared to all other IV construction methods.

\end{document}